\documentclass[aps,prl,twocolumn,showpacs,amsmath,amssymb]{revtex4}
\pdfoutput=1
\usepackage[latin2]{inputenc}
\usepackage{amsmath}
\usepackage{graphicx}
\usepackage{multirow}
\usepackage{dcolumn}

\newcommand{\ba}{\begin{eqnarray*}}
\newcommand{\ea}{\end{eqnarray*}}
\newcommand{\baa}{\begin{eqnarray}}
\newcommand{\eaa}{\end{eqnarray}}
\def\bar{\begin{array}}
\def\ear{\end{array}}
\def\LB{\left(}
\def\RB{\right)}

\def\pr{^{\prime}}

\def\Rr{\right>}
\def\Lb{\left|}

\def\uu{\uparrow}
\def\dd{\downarrow}
\def\pa{\uparrow\!\!\downarrow}
\def\s{\sigma}
\def\f{\frac}

\begin{document}

\title{Induced gauge potentials in reduced density matrix dynamics
} 

\author{Ryan Requist}
\email{rrequist@sissa.it}
\affiliation{
SISSA, via Bonomea 265, 34136 Trieste, Italy 
}

\date{\today}

\begin{abstract}

The combination of interactions and nonadiabaticity in many body systems is shown to induce magnetic gauge potentials in the equation of motion for the one-body reduced density matrix as well as the effective Schr\"odinger equation for the natural orbitals. The consequences of induced gauge geometry for charge and energy transfer are illustrated in the exact nonlinear dynamics of a three-site Hubbard ring ramped into a Floquet state by a time dependent circulating electric potential. Remarkably, the pumped charge flows against the driving in the strongly interacting regime, and the quasienergy level shift, which defines the work done on the system, can become negative.

\end{abstract}

\pacs{03.65.Vf, 
71.10.-w,   
72.10.-d,   
31.15.ee}

\maketitle

Gauge geometry is inherent to physical theories whose equations are phrased in terms of redundant variables.  The classical electromagnetic gauge potentials $(V,\mathbf{A})$ are redundant because infinitely many of them generate the same electromagnetic fields, yet they acquire a degree of observable significance in quantum physics through the Aharonov-Bohm effect \cite{aharonov1959}.  The phase factor \cite{wu1975} 
\begin{equation}
\mathrm{exp}\LB i\f{e}{\hbar} \oint A_{\mu} dx^{\mu} \RB,
\label{eq:exp}
\end{equation}
responsible for Aharonov-Bohm interference, is the fiber bundle holonomy of the 
connection $-i\f{e}{\hbar}A_{\mu}dx^{\mu}$ associated with local gauge invariance (gauge symmetry) \cite{fock1926,weyl1929a}.

Induced, as opposed to primitive, gauge geometries have gained attention only relatively recently.  Induced vector potentials were first found in the coupled equations for electronic and nuclear wavefunctions \cite{mead1979}, and the associated Aharonov-Bohm phase gives an alternative explanation for the sign change upon pseudorotation in triatomic molecules \cite{longuet-higgins1959}.  The discovery of geometric phase \cite{berry1984,simon1983,aharonov1987} established the geometric origin and observability of induced gauge geometries.  Induced vector potentials are in fact only the magnetic part of a general \textit{geometric electromagnetism} \cite{berry1989,jackiw1988,berry1993}, unifying induced electric and magnetic fields in a quantum geometric tensor \cite{provost1980,berry1989} over the space of slow variables.  The effective Hamiltonian for the slow variables generally also contains a geometric induced inertia \mbox{tensor \cite{goldhaber2005}}. 

Abelian and non-Abelian gauge geometries have found many applications in condensed matter physics, among which are adiabatic charge transport \cite{avron1988}, the theory of macroscopic polarization \cite{king-smith1993}, the anomalous velocity and other geometric effects of Bloch electrons \cite{chang1995,culcer2006,shindou2006,zak1989}, and the quantum Hall effect \cite{laughlin1981,avron1988,froehlich1993}.  Recent work has studied the Berry curvature in gradient expansions of the quantum kinetic equations of Fermi liquids \cite{shindou2006,wong2011}.  Another line of research aims to simulate condensed matter phases by realizing artificial gauge potentials for trapped ultracold neutral atoms \cite{dum1996,lin2009,dalibard2011}.  

The above gauge geometries were formulated for noninteracting systems or at the mean field level.  Although induced gauge potentials and geometric phases are equally valid for interacting systems, they are difficult to compute if the complexity of the many body wave function scales exponentially with the number of particles.  For this reason, it is desirable to identify geometric structures at the finer level of $n$-body reduced density matrices (rdms), defined through the partial trace $\rho_n = \binom{N}{n} \mathrm{Tr}_{n+1\ldots N} \rho$.  Reduced geo\-metric phases for $n$-body rdms are one example \cite{requist2012}. 

The purpose of this Letter is to point out the \mbox{existence} of induced gauge geometries in the equations of motion for $n$-body rdms $\rho_1$, $\rho_2$, $\rho_3$, $\ldots$, which are organized into a chain-like structure called the quantum Bogoliubov-Born-Green-Kirkwood-Yvon (BBGKY) hierarchy.  These multi\-farious gauge structures are associated with the gauge freedom induced by separating the rdm variables into a hierarchy of levels $\rho_1\backslash\rho_2\backslash\rho_3\backslash \ldots$, in agreement with Berry's notion that induced gauge geometries result from the division of a composite system into two parts \cite{berry1989}.  For example, at the first level of the hierarchy, the marginal density $\rho_1$ acts like the nuclear wavefunction $\Phi(R)$ in the Born-Oppenheimer approximation, while the density $\rho_2$ (conditional on $\rho_1$) acts like the electronic factor $\psi_R(r)$.  In the simplest (Abelian) case, the gauge variables are $U(1)$ phases corresponding to the unitary transformation $V^{\dag} \rho_2 \,V$, where $V= e^{i\phi_k n_k}$ and $n_k$ is the number operator for a $\rho_1$ eigenstate.  Our results suggest an extension of geometric electromagnetism to many body systems and establish the BBGKY hierarchy as a framework for applying differential geometry to many body dynamics.

The physical effects of induced gauge potentials are exemplified here in a three-site Hubbard ring ramped into a Floquet state by a circulating potential well.  Induced gauge potentials mediate energy transfer through the electromotive force implied by dynamical variations of the induced magnetic flux (Faraday's law).  
We find an intriguing many body effect whereby the pumped charge flows backwards against the driving fields when Hubbard interactions are sufficiently strong.  The work done on the system by the driving fields during the adiabatic ramping is given by the quasienergy level shift, and surprisingly, it can become negative.

\begin{figure}[t!]
\begin{tabular}{@{}cl@{}}
\includegraphics[width=0.19\columnwidth]{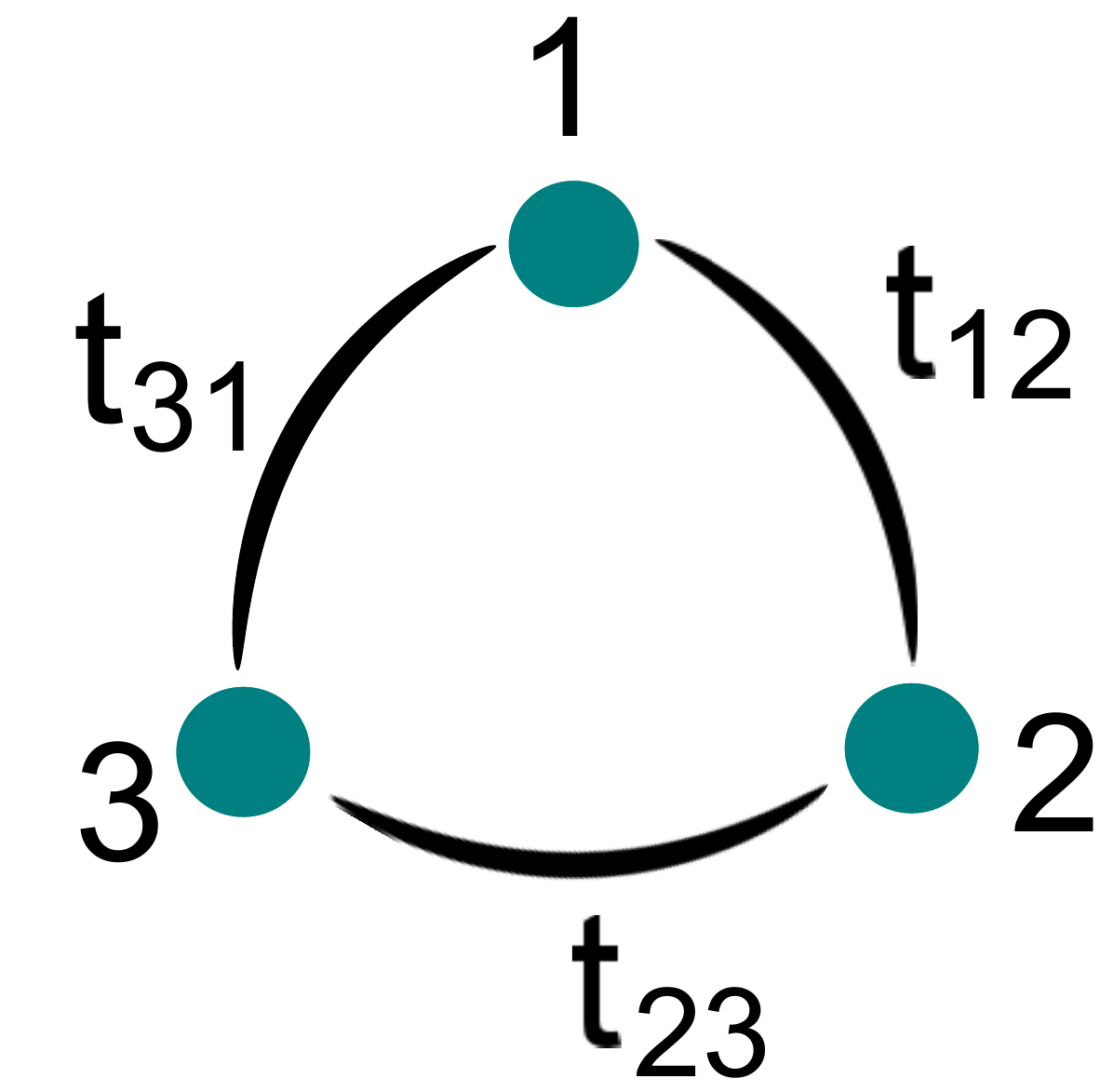} & \multirow{2}{*}[-0.18cm]{\includegraphics[width=0.48\columnwidth]{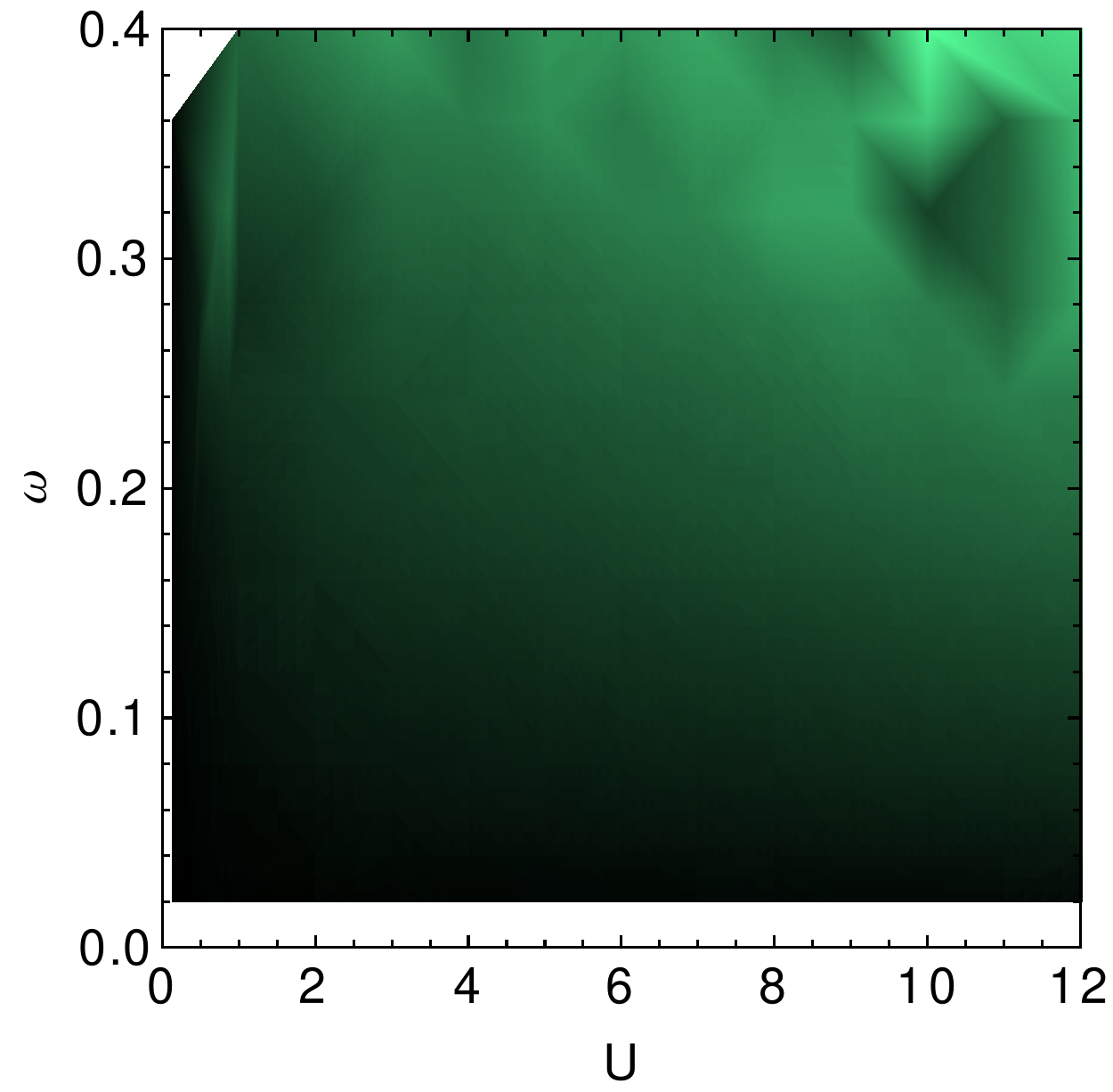}} \\
\includegraphics[width=0.44\columnwidth]{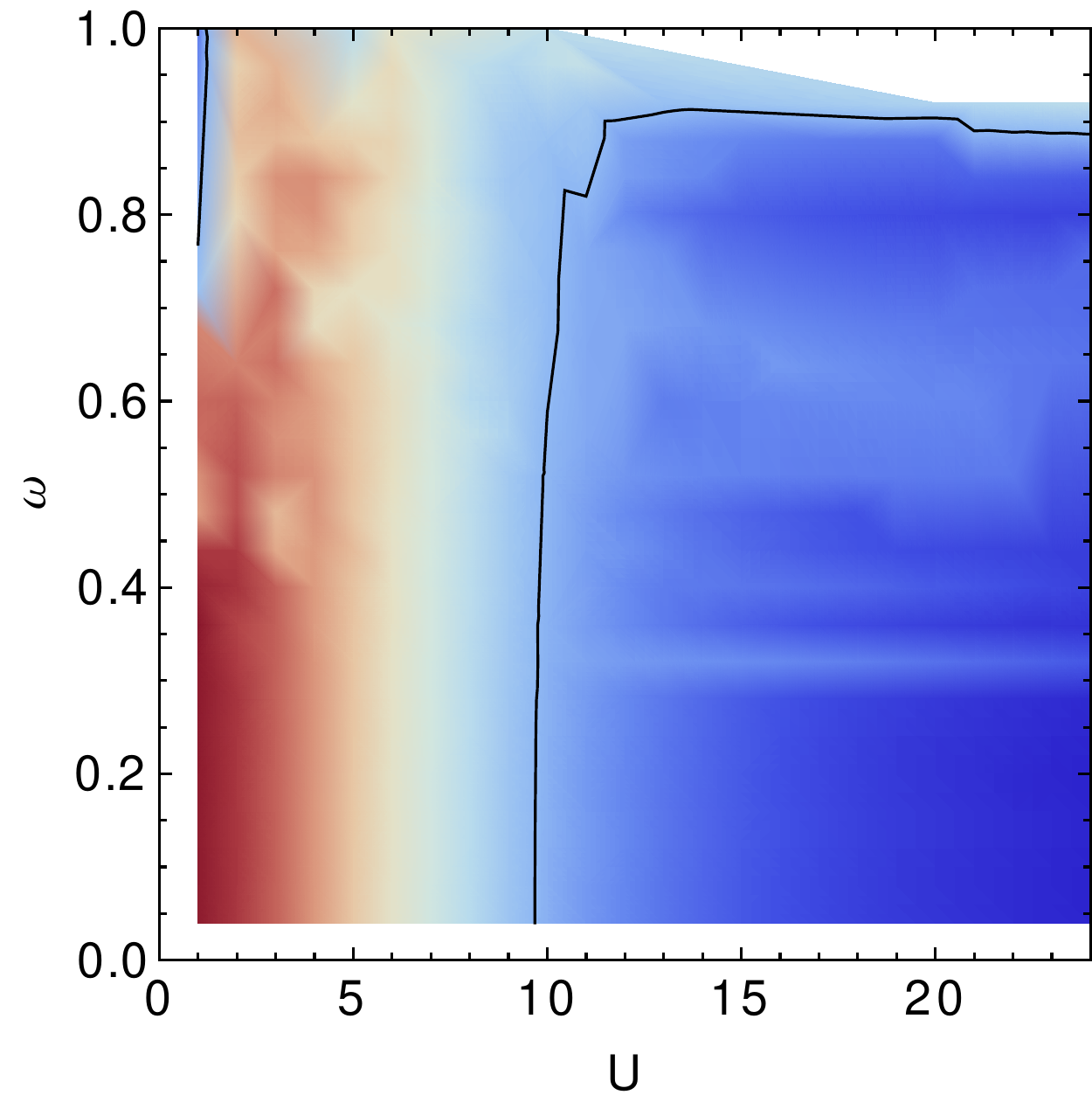}\hspace{0.00cm} \includegraphics[width=0.045\columnwidth]{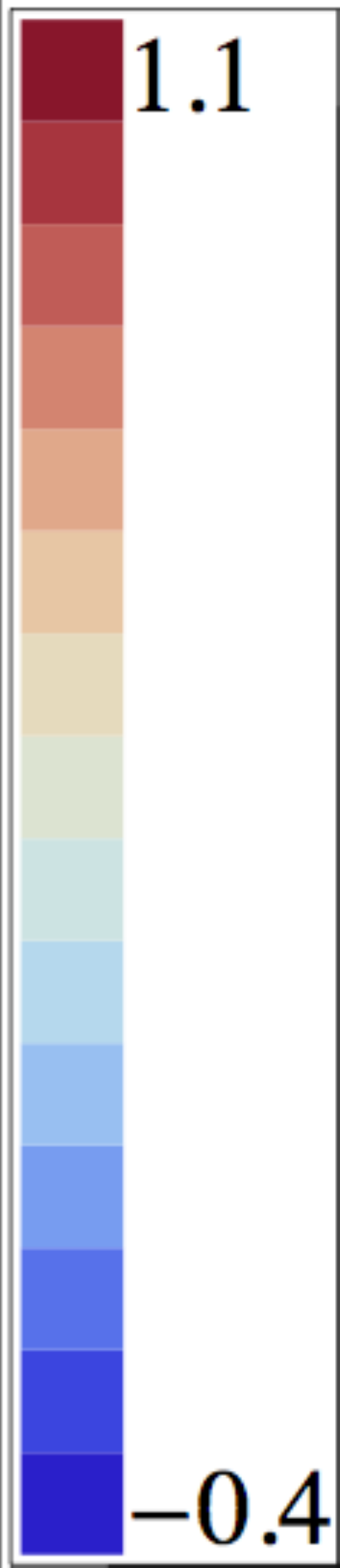} & 
\end{tabular}
\caption{(a) Three-site Hubbard ring, (b) pumped charge $Q$ and (c) ratio of the real to imaginary part of the time averaged gauge-invariant loop quantity $u_{12} u_{23} u_{31}$ as a function of $(U,\omega_0)$ for ramping speed $\alpha=0.11$, well depth $\epsilon=4$, $U=7$ and initial phase $\phi_0=2\pi/3$. Scale: black=0, light green=0.4.}
\label{fig:Phiu} 
\end{figure} 

Our starting point is the first equation of the BBGKY hierarchy, the dynamical equation for the operator $\rho_1$,
\begin{align}
i \partial_t \rho_1 = \left[ v, \rho_1 \right] + iu {,}
\label{eq:eom:rho1}
\end{align}
where $v = \mathbf{p}^2/2m+V(\mathbf{r}t)$ and the Hermitian operator $u$ is nonlocal in coordinate/spin space 
\begin{align}
\langle \mathbf{r} |u| \mathbf{r}\pr\rangle = \f{2}{i} \int d^3z \left[ \f{e^2}{|\mathbf{r}-\mathbf{z}|} - \f{e^2}{|\mathbf{r}\pr-\mathbf{z}|} \right] \rho_2(\mathbf{r} \mathbf{z}|\mathbf{r}\pr \mathbf{z}), \label{eq:u}
\end{align}
suppressing spin indices.  Although $\langle \mathbf{r} |u| \mathbf{r}\pr\rangle$ is not invariant to local gauge transformations, it nevertheless contains gauge invariant information.  This is easily seen in the context of lattice models, where
vector potentials are represented by Peierls phases on the links between sites, i.e.~$t_{ij}\rightarrow t_{ij} e^{i A_{ij}}$.  A given lattice Hamiltonian has magnetic fields if and only if the flux through a plaquette, a gauge-invariant quantity, is nonzero,
\begin{align}
\Phi = \sum_{n=1}^N A_{n,n+1} \neq 0, \label{eq:Phi}
\end{align}
where the sum runs over a circuit of sites $n=1,2,\ldots N$ and site $N+1$ is the same as site 1.  The phase factor $e^{i\Phi}$ is analogous to the Wilson loop phase factor in lattice gauge theory \cite{wilson1974}.  For simplicity, we restrict our attention to $U(1)$ lattice gauge theory to avoid complications associated with path ordering.

The minimal model realizing nontrivial induced gauge potentials is a three-site Hubbard ring (Fig.~\ref{fig:Phiu}b) with the Hamiltonian 
\begin{align}
H &= - \sum_{i,\s} \LB t_{i,i+1} \; c_{i\s}^{\dag} c_{i+1\s} + H.c. \RB + \sum_{i,\s} \epsilon_i(t) \, \hat{n}_{i\s} \nonumber \\
&\quad+ U (\hat{n}_{1\uu} \hat{n}_{1\dd} + \hat{n}_{2\uu} \hat{n}_{2\dd} + \hat{n}_{3\uu} \hat{n}_{3\dd}), \label{eq:H}
\end{align}
which describes electrons that hop with amplitudes $t_{i,i+1}$ among three sites (we set $t_{i,i+1}=1$).  Coulomb interactions are approximated by a local on-site Hubbard form $\mathcal{U} = U (\hat{n}_{1\uu} \hat{n}_{1\dd} + \hat{n}_{2\uu} \hat{n}_{2\dd} + \hat{n}_{3\uu} \hat{n}_{3\dd})$.  The sites can represent atomic orbitals, quantum dots, impurities, etc.  For example, this three-site Hubbard ring was used to model the quantum electric dipole moment (a geo\-metric effect) of triatomic molecules and triple quantum dots \cite{requist2005b}.  Our main result is that the reduced equations of motion contain \textit{magnetic} gauge potentials even though Eq.~(\ref{eq:H}) has no external magnetic fields, since $t_{i,i+1}$ are real and $\epsilon_i$ represent purely electric driving.  We demonstrate the existence of induced gauge potentials in two quantities: (I) the operator $u$ in Eq.~(\ref{eq:eom:rho1}) and (II) the effective Hamiltonian $h$ for the natural orbitals (defined below).

\textit{Gauge geometry type I} --- The quantity
\begin{align}
\Phi_u = \mathrm{Arg} \: u_{12} u_{23} u_{31} - \f{\pi}{2}
\end{align}
with $u_{jk} = (2U/i)\sum_l (\delta_{jl}-\delta_{kl})\rho_{2,jlkl}$ is gauge invariant like $\Phi$ in Eq.~(\ref{eq:Phi}), cf.~$\Phi =\mathrm{Arg} \: t_{12} t_{23} t_{31}$.  If $\Phi_u \neq 0$, the operator $u$ contains a magnetic-type flux, implying nontrivial gauge geometry associated with closed loops in coordinate space.  Like the bare magnetic flux present in $H$ if $\Phi\neq 0$, it cannot be removed by any gauge transformation.  Figure 1c shows the ratio of the real to imaginary part of $\int_t^{t+T/6} u_{12} u_{23} u_{31}ds$ at long times for a driven ring with two electrons in a spin singlet.  This proves that $u_{12} u_{23} u_{31}$ is not purely imaginary, and hence $\Phi_u\neq 0$.  The implied gauge geometry is due to the cooperation of interactions and nonadiabaticity.  This is intuitively clear since in the noninteracting limit, $u\rightarrow 0$ in Eq.~(\ref{eq:eom:rho1}) and $v$ has no magnetic fields by assumption.  At the same time, the nonadiabatic transitions between instantaneous eigenstates responsible for inducing the magnetic flux vanish in the limit $\omega_0\rightarrow 0$ (see \cite{sm} for the derivation of an adiabatic effective many body Hamiltonian). Figure 1c provides numerical support for these conclusions, showing that the time averaged real part of $u_{12} u_{23} u_{31}$ vanishes faster than imaginary part in both the noninteracting and adiabatic limits.  The $\epsilon_i$ driving is parameterized by $\epsilon_1 = V_3/2 + V_8/2\sqrt{3}$, $\epsilon_2 = - V_3/2 + V_8/2\sqrt{3}$, and $\epsilon_3 = -V_8/\sqrt{3}$ with (notations explained in \cite{sm}) 
\begin{align}
V_3 &= -\sqrt{3} \epsilon \:\sin\Big[ \phi(t) +\phi_0 \Big] \nonumber \\
V_8 &= +\sqrt{3} \epsilon \:\cos\Big[ \phi(t) +\phi_0 \Big],
\label{eq:driving}
\end{align}
where $\phi(t)=\omega_0(t + \f{1}{\alpha} \log 2\cosh \alpha t)/2$, corresponding to the frequency $\omega(t) = d\phi/dt = \omega_0 (1+\tanh \alpha t)/2$.  This describes a potential well, localized at site 1 at $t=-\infty$ if $\phi_0=2\pi/3$, which slowly increases its rate of circulation around the ring ultimately reaching a constant rotational speed $\omega_0$.  In all cases, we choose the initial state to be the ground state.  The ground state energy depends on $\phi_0$.  For small $U$, the most stable ground state occurs for $\phi_0=0 \!\!\mod 2\pi/3$, since for that value both electrons can lower their energy by occupying the potential well; 
\begin{figure}[htb!]
\includegraphics[width=\columnwidth]{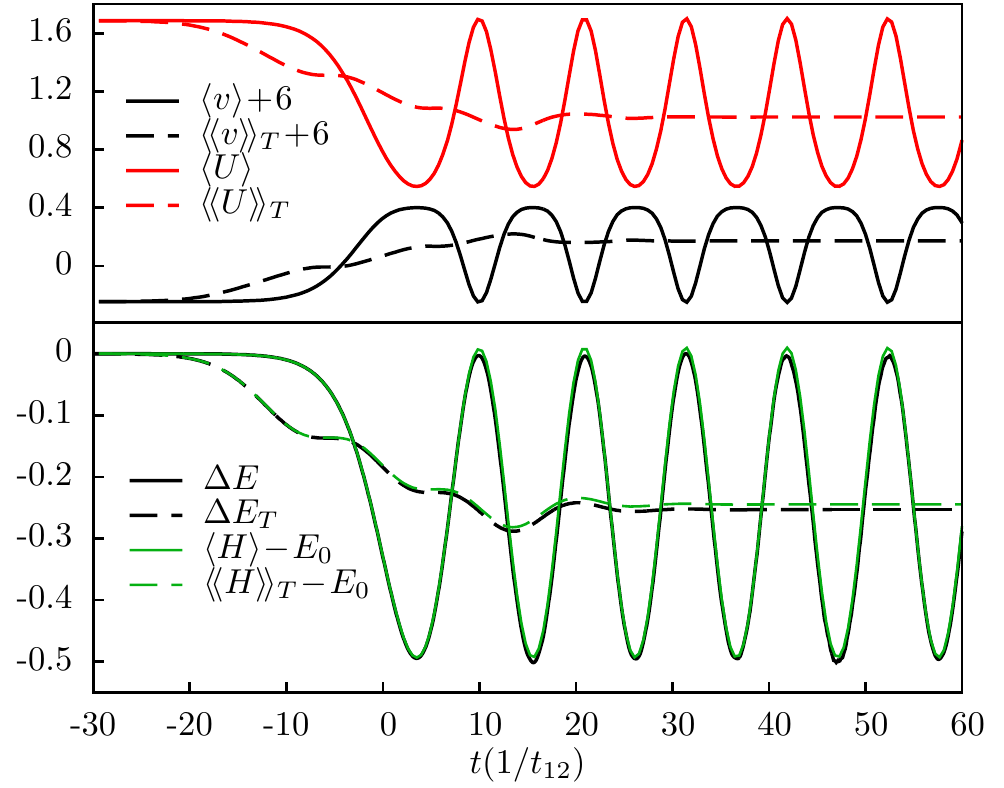}
\caption{(a) Instantaneous one-body $\langle v \rangle$ and interaction $\langle \mathcal{U} \rangle$ energies, (b) the quasienergy level shift $\Delta E$ and instantaneous energy $\langle H\rangle$; all shown with running time averages. Same parameters as Fig.~\ref{fig:Phiu}.}
\label{fig:energy12} 
\end{figure} 
however, as $U$ is increased and it becomes unfavorable for both electrons to occupy the same site, the ground state undergoes a transition to a delocalized state for which the most stable value of $\phi_0$ is $\pi \!\!\mod 2\pi/3$.  For $\phi_0=\pi$, the potential well is halfway between sites 1 and 2, so they are initially degenerate.  Cyclic driving protocols similar to Eq.~(\ref{eq:driving}) have been realized in trapped Bose-Einstein condensates \cite{wright2013a,arwas2013} and could be implemented in triple quantum dots (see Ref.~\cite{hsieh2012} and references cited therein). 

The operator $u$ describes how interactions affect the dynamics of $\rho_1$ by mediating energy transfer between collective variables and internal interaction energy.  To see this, consider the time derivative of the one-body energy 
\begin{align}
\f{d\langle v\rangle}{dt} &= \mathrm{Tr}\LB\rho_1 \partial_t v\RB + \mathrm{Tr}(u v)  \nonumber \\
&= \sum_i n_i \dot{\epsilon}_i + (4 + \epsilon^3 \cos3(\phi+\phi_0)) |u_{12}u_{23}u_{31}| \sin\Phi_u \nonumber
\end{align}
The first term is the power applied to the whole system by the external driving, and the second term 
is the power applied on the one-body variables by two-body interactions; the latter is \textit{modulated} by the flux $\Phi_u$ and vanishes when $\Phi_u=0$.  Figure~\ref{fig:energy12}a illustrates energy exchange between $\langle v \rangle$ and $\langle \mathcal{U} \rangle$; also shown are the running time averages, e.g.~$\langle\!\langle v(t) \rangle\!\rangle_T = (1/T) \int_{t-T/2}^{t+T/2} \langle v(s) \rangle ds$; $T=2\pi/\omega_0$.

The Hamiltonian in (\ref{eq:H}) becomes $T$-periodic as $t\rightarrow \infty$. For sufficiently small $\alpha$ the system evolves adiabatically from the ground state $|\psi_0\rangle$ to a Floquet state $|\psi(t)\rangle$ \cite{sm}.  The wavefunction can be split into a factor $|\xi(t)\rangle$ which becomes periodic in the steady state and an overall phase factor \cite{langhoff1972,requist2009}
\begin{align}
|\Psi(t)\rangle = e^{-i\int_{-\infty}^t \Omega(s)ds} |\xi(t)\rangle. \label{eq:split}
\end{align}
This factorization is not unique, and for convenience we have chosen $|\xi(t)\rangle = e^{-i\mathrm{Arg}\langle \psi_0|\psi(t)\rangle} |\psi(t)\rangle$.  This choice gives an oscillatory $\Omega(t) = i \partial_t \log \langle \psi_0|\psi(t)\rangle$ as shown in Fig.~\ref{fig:energy12}b, but the running time average $\Omega_T(t)$ approaches a constant asymptotic quasienergy 
\begin{align}
\Omega = \f{\langle \xi | H(t) - i \partial_t | \xi \rangle}{\langle \xi | \xi \rangle}.\label{eq:Omega}
\end{align}
The quasienergy of a Floquet state is only defined modulo $\omega_0$, implying that the set of quasienergies have a Brillioun zone structure $[0,\omega_0]$, $[\omega_0,2\omega_0]\ldots\:$ 
It is possible to make a different gauge choice for $|\xi(t)\rangle$ such that the path $\Omega_T(t)$ approaches $\Omega + n\omega_0$ for any $n$.  The integer $n$, which corresponds to a winding number of $|\xi(t)\rangle$, is a topological quantity in the sense that any two paths that end in different zones at $t=\infty$ cannot be smoothly transformed into each other.  Nevertheless, there is a class of gauge choices for which $\Omega_T(t)$ remains close to the adiabatically continued quasienergy $\Omega_0(t)$ of the instantaneous Floquet state $|\xi_0(t)\rangle$, thereby defining a unique $\Omega$ \cite{sm}. 

The constant asymptotic level shift $\Delta E=\Omega-E_0$ represents the work done on the system in the course of ramping on the perturbation.  Evidence of that work is done is visible in the running time average $\langle\!\langle H(t) \rangle\!\rangle_T$ shown in Fig.~\ref{fig:energy12}b, which changes from $E_0$ to a constant value close to $\Omega$. The persistent oscillations in $\langle H(t) \rangle$ represent the continuous exchange of energy, back and forth, between the system and its environment, i.e.~the collection of charges, currents and fields responsible for producing the given electric driving (we neglect the small associated magnetic fields).  In order for $\Delta E$ to provide a consistent definition of work, it is imperative to keep track of its zone. In second-order perturbation theory, $\Delta E$ as observable in the ac Stark shift \cite{langhoff1972}. 
%
%
%

\begin{figure}[tbh!]
\includegraphics[width=\columnwidth]{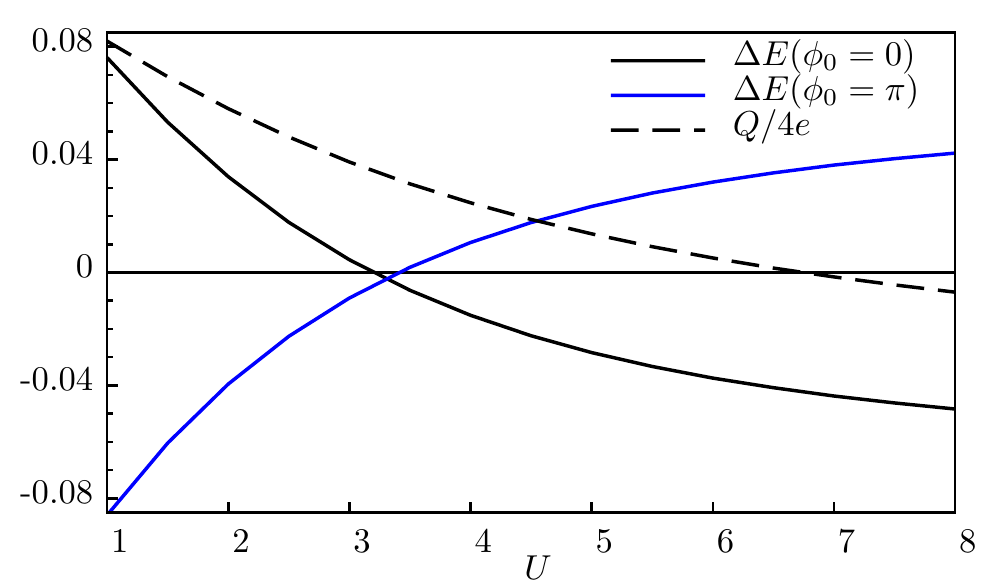}
\caption{Quasienergy level shift $\Delta E$ and pumped charge $Q$ as a function of $U$ for $\alpha=0.125$, $\omega_0=0.3$ and $\epsilon=4$.}
\label{fig:QDeltaE} 
\end{figure}

Figure~\ref{fig:QDeltaE} shows how $\Delta E$ changes as a function of $U$.  The quasienergy level shift can assume negative values.  First, consider small $U$.  If $\phi_0=0$, the system starts in the stable ground state and $\Delta E$ is positive because the driving does work on the system by increasing its time-averaged kinetic plus potential energy $\langle\!\langle v \rangle\!\rangle_T$.  On the other hand, if $\phi_0=\pi$, the system starts in the unstable ground state and $\Delta E$ is negative because the system lowers $\langle\!\langle v \rangle\!\rangle_T$ by starting to rotate.  Now, consider large $U$.  The situation is reversed, and $\Delta E$ is negative for $\phi_0=0$ and positive for $\phi_0=\pi$.  For $\phi_0=0$, the work done on the system is negative because the stabilization of the interaction energy $\langle \mathcal{U} \rangle$ more than compensates for the increase in $\langle v\rangle$, as shown in Fig.~\ref{fig:energy12}a.  Another physical mechanism that stabilizes the rotating state is geometric phase.  The geometric phase contribution to $\Omega$ is $-i \f{1}{T}\int_t^{t+T} \langle \xi |\partial_s \xi \rangle ds$, and since it is negative, it stabilizes the rotating state.  This stabilization can be seen in the small offset of $\Omega_T(t)$ from $\langle\!\langle H(t) \rangle\!\rangle_T$ in Fig.~\ref{fig:energy12}b.  Remarkably, there is a small region near $U\approx 3.3$ where the system gives up energy by adopting a rotating state for \textit{any} initial phase $\phi_0$.

\textit{Gauge geometry type II} --- A second type of induced gauge potential appears in the Hamiltonian $h$ governing the dynamics of single-particle states $|\psi_k\rangle$ defined as
\begin{align}
|\psi_k\rangle = \sqrt{n_k} e^{-i\zeta_k} |\phi_k\rangle, \label{eq:psik}
\end{align}
where $|\phi_k\rangle$ is a natural orbital (eigenstate of $\rho_1$), $n_k$ is the occupation number and $\zeta_k$ is the phase conjugate to $n_k$ \cite{requist2011}.  
The states in Eq.~(\ref{eq:psik}) were introduced in Ref.~\onlinecite{giesbertz2010}, except without the factor $\sqrt{n_k}$.  The factor $\sqrt{n_k}$ was added in Ref.~\onlinecite{requist2012} for a geometric reason, namely $i\langle \psi_k | d\psi_k\rangle$ constitutes a connection one-form whose holonomy is a reduced geometric phase.  For the two-electron system considered here, the set $\{|\psi_k\rangle\}$ contains all of the degrees of freedom of $|\Psi\rangle$ in a compact form, which is apparent from the expression $|\Psi\rangle = \sum_k \sqrt{n_k/2} e^{-i\mu-i2\zeta_k} |\phi_k\phi_k\rangle$.  Since the induced magnetic flux enters $h$ exactly as an external magnetic flux does, it has a more straightforward interpretation than the type I induced flux. 

\begin{figure}[th!]
\includegraphics[width=\columnwidth]{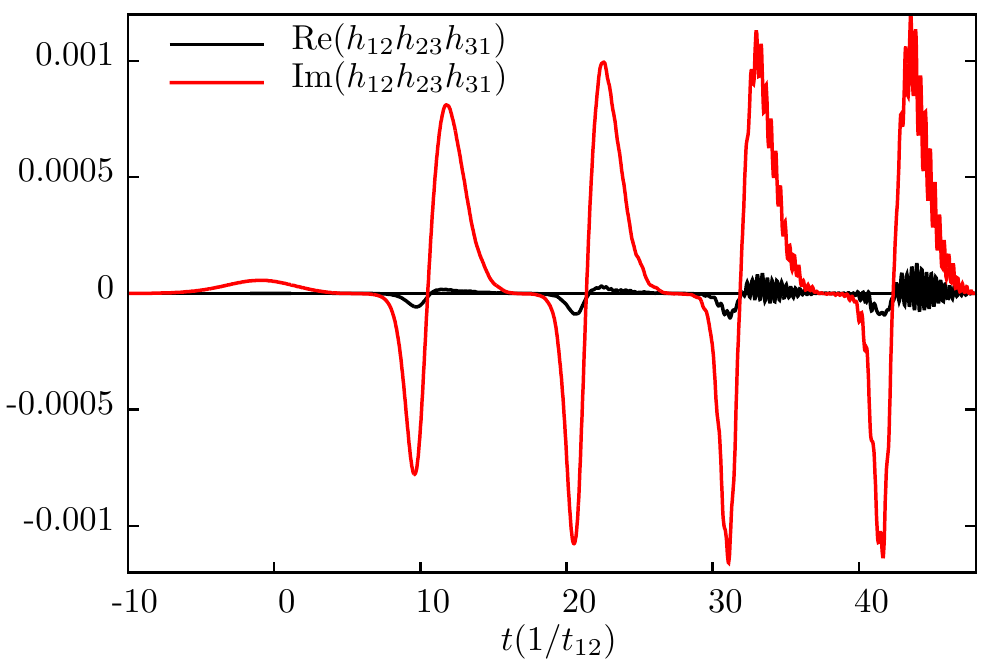}
\caption{Real and imaginary parts of the gauge invariant loop quantity $h_{12} h_{23} h_{31}$ for the same parameters as Fig~\ref{fig:Phiu}.}
\label{fig:Phih} 
\end{figure}

The effective Schr\"odinger equation for the $|\psi_k\rangle$ is 
\begin{align}
i\partial_t |\psi_k\rangle = h |\psi_k \rangle, \label{eq:schroedinger}
\end{align}
where $h$ must be non-Hermitian since it changes the modulus of $|\psi_k\rangle$.  The criterion for $h$ to have induced magnetic fields is $\Phi_h = \mathrm{Arg}\: h_{12} h_{23} h_{31} \neq 0$.  Figure~\ref{fig:Phih} shows the real and imaginary parts of $h_{12} h_{23} h_{31}$. The elements of $h$ are very strongly renormalized with respect to the given $v$. The renormalization of the hopping and on-site elements $h_{ij}$ can be understood along the lines of the renormalization in the Gutzwiller approximation.

That $h$ must contain magnetic fields is not obvious.  If all the reduced system had to do was pump charge, electric fields would be sufficient.  However, $h$ must also reproduce the dynamics of all $|\psi_k\rangle$, including their individual dynamical and geometric phases, and that would not be possible without induced magnetic fields.  A similar situation would occur if time dependent current density functional theory \cite{ghosh1988,vignale2004} were applied to the present problem because the noninteracting Kohn-Sham system would contain an induced vector potential $\mathbf{A}_{xc}$ even in the absence of externally applied magnetic fields. 

%
%
%

Including the $\zeta_k$ phases in the definition of the $|\psi_k\rangle$ makes them properly gauge invariant and allows us to define individual quasienergies $\Omega_k$ by applying the same factorization as in Eq.~(\ref{eq:split}).  In order for the two-body state $|\xi\rangle$ to be periodic in the long-time regime, we must have $\Omega_k = \Omega\, \mathrm{mod}\, \omega_0$.  Nevertheless, $\Omega_k(t)$ can have quite different time profiles within one period, and the phase variables $\zeta_k$ display nontrivial winding numbers.

The pumped charge $Q$ is a decreasing function of $U$, as shown in Figs.~1b and \ref{fig:QDeltaE}, because the electric driving fields become less effective for large $U$.  Electric fields are effective in pumping charge only insofar as there are imbalances between the site occupations, and for large $U$ the amplitude of the periodic oscillations in the site occupancies is suppressed by strong two-body correlations which inhibit double occupancy.  

Remarkably, the pumped charge becomes negative for large $U$.  This is a many body effect related to the fact that for large $U$ the site occupations are pinned to 1, giving two singly occupied sites and one empty site.  Although charge flows with the driving along the links nearest to the potential well, there is an even larger backwards current along the link opposite to the well.  The pumped charge is related to the quasienergy according to
\begin{align}
\f{Q}{e} = -\f{\partial \Omega T}{\partial \Phi}, \label{eq:Q}
\end{align}
which is similar to a formula for Cooper pair pumping in superconducting circuits \cite{russomanno2011}.  Equation~(\ref{eq:Q}) is related to the stationarity of the quasienergy $\Omega[J_T]$ \cite{sm}. The reduced geometric phases $\int_t^{t+T} i\langle \psi_k |\partial_s \psi_k\rangle ds$ contribute to the pumped charge since their sum gives the geometric contribution to the quasienergy. 
%
Apart from the coupling to external electric fields, our model is a closed system.  The effect of dissipation on the pumped charge in a noninteracting three-site ring coupled to a bath of harmonic oscillators has been studied \cite{pellegrini2011}.  

In summary, we identified two types of induced gauge geometry resulting from the conjunction of interactions and nonadiabaticity in many body systems.  The implications of the associated effective magnetic fields for charge and energy transfer were illustrated in a driven three-site Hubbard ring; the predicted phenomena are potentially observable in triple quantum dots or ultracold atoms.

\begin{acknowledgments}
Early stages of the research were supported by the Deutsche Forschungsgemeinshaft (Grant No.PA516/7-1) and later stages by PRIN/COFIN 2010LLKJBX 004 and 2010LLKJBX 007, Sinergia CRSII2136287/1 as well as ERC Advanced Grant 320796 -- MODPHYSFRICT. 
\end{acknowledgments}


\pagebreak
\widetext

\setcounter{equation}{0}
\setcounter{figure}{0}
\setcounter{table}{0}
\setcounter{page}{1}
\makeatletter

\renewcommand{\thefigure}{S\arabic{figure}}
\renewcommand{\thetable}{S\arabic{table}}
\renewcommand{\theequation}{S\arabic{equation}}
\renewcommand{\bibnumfmt}[1]{[S#1]}
\renewcommand{\citenumfont}[1]{S#1}


\begin{center}
{\Large \bf Supplemental Material}
\end{center}
\bigskip

\noindent {\bf S1. Lie algebra parameterization, angular parameterization and a generalized Bloch equation}
\medskip

The analysis of the three-site Hubbard model involves Hermitian $3\times 3$ matrices, which are naturally parameterized using the $\mathfrak{su}(3)$ Lie algebra.  For example, the spin-summed one-body rdm $\rho_{1,ij} = \sum_{\sigma} \rho_{1,i\sigma j\sigma}$ can be expanded as 
\begin{align}
\rho_1 = \vec{\rho}_1 \cdot \vec{\nu},
\end{align}
where $\vec{\nu}$ is a nine-component vector whose first eight elements are the Gell-Mann matrices and whose ninth element is the identity matrix.  Similarly, the one-body terms of the Hamiltonian in Eq.~(5) can be expressed as 
\begin{align}
v = \f{1}{2} \vec{V} \cdot \hat{\vec{\nu}},
\label{eq:one-body}
\end{align}
where $\hat{\vec{\nu}}$ is the vector of operators $\hat{\nu}_k = \sum_{\mu\nu\sigma} c_{\mu\sigma}^{\dag} \nu_{k,\mu\nu} c_{\nu\sigma}$ and the elements of  
$\vec{V}=\mathrm{Tr}(\hat{H}\hat{\vec{\nu}})$ are
\begin{align}
V_1 &= -2 \,\mathrm{Re} \; t_{12} = -2  & V_4 &= -2 \,\mathrm{Re} \; t_{31} = -2 & V_6 &= -2 \,\mathrm{Re} \; t_{23} = -2 \nonumber \\
V_2 &= -2 \,\mathrm{Im} \; t_{12} = 0  & V_5 &= -2 \,\mathrm{Im} \; t_{31} = 0 & V_7 &= -2 \,\mathrm{Im} \; t_{23} = 0\nonumber \\
V_3 &= \epsilon_1-\epsilon_2 & V_8 &= \f{1}{\sqrt{3}} (\epsilon_1+\epsilon_2-2\epsilon_3) & V_9 &= \sqrt{\f{2}{3}} (\epsilon_1+\epsilon_2+\epsilon_3).  \label{eq:V}
\end{align}
The on-site energies $\epsilon_i$ depend only on the variables $(V_3,V_8,V_9)$.  In the main text, the driving is chosen to be
\begin{align}
V_3 &= -\sqrt{3} \epsilon \:\sin\Big[ \phi(t) +\phi_0 \Big] \nonumber \\
V_8 &= +\sqrt{3} \epsilon \:\cos\Big[ \phi(t) +\phi_0 \Big] \label{eq:driving:s}
\end{align}
and, without loss of generality, the spatial constant $V_9$ is set to zero.  The factor $\sqrt{3}$ is introduced to normalize the $\epsilon_i$, so for example $(\epsilon_1,\epsilon_2,\epsilon_3)$ equals $\epsilon(-1,\f{1}{2},\f{1}{2})$ if $\phi+\phi_0=\f{2\pi}{3}$ and $\epsilon(-\f{1}{2},-\f{1}{2},1)$ if $\phi+\phi_0=\pi$.

Since the number of electrons is conserved, the occupation numbers $n_k$ (eigenvalues of $\rho_1$) can be parameterized as
\begin{align}
n_a &= \f{2}{3} + \f{A}{3} + B  \nonumber \\
n_b &= \f{2}{3} + \f{A}{3} - B \nonumber \\
n_c &= \f{2}{3} - \f{2A}{3},
\end{align}
where $A=(n_a+n_b-2n_c)/2$ and $B=(n_a-n_b)/2$ satisfy the inequality constraints $0\leq B \leq A\leq 1$ due to the Pauli principle and because we choose $n_c\leq n_b\leq n_a$.

The natural orbitals $|\phi_k\rangle$ are parameterized in terms of 6 angle variables $(\theta_1, \theta_2, \theta_3, \varphi_1, \varphi_2, \varphi_3)$ as follows \cite{s-bronzan1988}
\begin{align}
\phi_a &= \LB \bar{r} \cos\theta_1 \cos\theta_2 \, e^{-i(\varphi_1-\varphi_2)/2} \\ 
\cos\theta_1 \sin\theta_2 \, e^{-i(\varphi_1+\varphi_2)/2}  \\ 
\sin\theta_1 \, e^{+i(\varphi_1+\varphi_2)/2} \ear \RB \nonumber \\[0.4cm]
\phi_b &= \LB \bar{c} -\sin\theta_1 \cos\theta_2 \sin\theta_3 \, e^{-i(\varphi_1+\varphi_3)/2} - \sin\theta_2 \cos\theta_3 \, e^{+i(\varphi_1+\varphi_3)/2} \, e^{+i\varphi_2} \\
+\cos\theta_2 \cos\theta_3 \, e^{+i(\varphi_1+\varphi_3)/2} - \sin\theta_1 \sin\theta_2 \sin\theta_3 \, e^{-i(\varphi_1+\varphi_3)/2} \, e^{-i\varphi_2} \\
+\cos\theta_1 \sin\theta_3 \, e^{+i(\varphi_1-\varphi_3)/2} \ear \RB \nonumber \\[0.4cm]
\phi_c &= \LB \bar{c} -\sin\theta_1 \cos\theta_2 \cos\theta_3 \, e^{-i(\varphi_1+\varphi_3)/2} + \sin\theta_2 \sin\theta_3  \, e^{+i(\varphi_1+\varphi_3)/2} \, e^{+i\varphi_2} \\ 
-\cos\theta_2 \sin\theta_3 \, e^{+i(\varphi_1+\varphi_3)/2} - \sin\theta_1 \sin\theta_2 \cos\theta_3 \, e^{-i(\varphi_1+\varphi_3)/2} \, e^{-i\varphi_2} \\
+\cos\theta_1 \cos\theta_3 \, e^{+i(\varphi_1-\varphi_3)/2} \ear \RB.
\label{eq:phik}
\end{align}
The variables $(\theta_3,\varphi_3)$ describe beyond mean field dynamics because they are not present in the most strongly occupied orbital $|\phi_a\rangle$, which is the only occupied orbital in a mean field-like theory.  To investigate the structure of the $(\theta_3,\varphi_3)$ subspace, consider the unitary transformation from the site basis of Eq.~(\ref{eq:phik}) to the basis $(\phi_a, \phi_u, \phi_v)$, where 
\begin{align}
\phi_u &= \LB \bar{c} -\sin\theta_2 \:e^{+i \varphi_2/2} \\ +\cos\theta_2 \:e^{-i \varphi_2/2}\\ 0 \ear \RB,  \qquad 
\phi_v = \LB \bar{r} -\sin\theta_1 \cos\theta_2 \:e^{-i (\varphi_1 - \varphi_2)/2} \\ -\sin\theta_1 \sin\theta_2 \:e^{-i (\varphi_1 + \varphi_2)/2} \\ \cos\theta_1 \:e^{+i (\varphi_1 + \varphi_2)/2} \ear \RB
\end{align}
are two states orthogonal to $\phi_a$.  In this basis $\rho_1$ is block diagonal
\begin{align*}
\rho_1 &= \LB \bar{ccc} n_a & 0 & 0 \\ 0 & \f{1}{2} (n_b+n_c) & 0 \\ 0 & 0 & \f{1}{2} (n_b+n_c) \ear \RB + \f{1}{2} (n_b-n_c) \LB \bar{ccc} 0 & 0 & 0 \\ 0 & \cos2\theta_3 & \sin2\theta_3 \; e^{-i2\varphi_3} \\ 
0 & \sin2\theta_3 \; e^{+i2\varphi_3}  & -\cos2\theta_3 \ear \RB,
\end{align*}
and we see that $(\theta_3,\varphi_3)$ parameterize the orbit of an SU(2) subgroup of SU(3) acting on $\rho_1$.

\begin{figure}[t!]
\includegraphics[width=0.55\columnwidth]{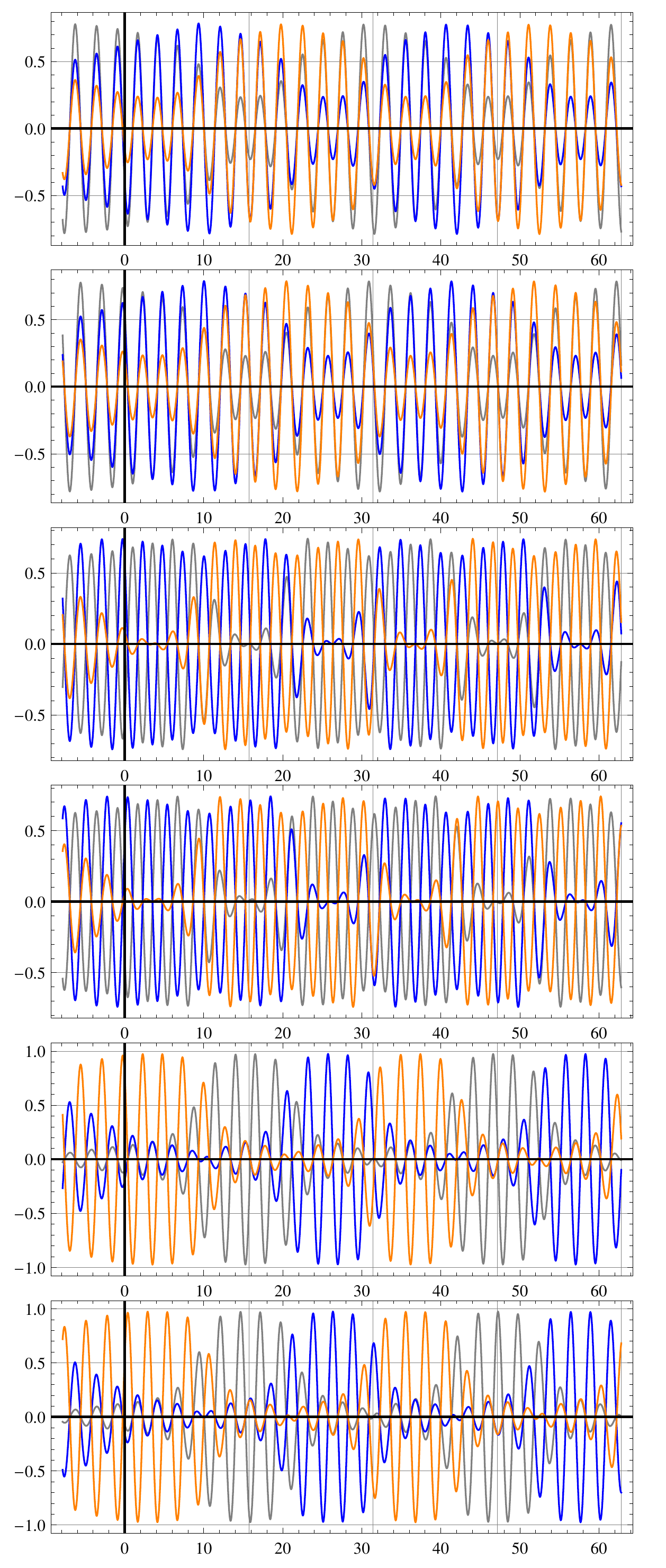}
\caption{Dynamics of the real and imaginary parts of the components (1=gray, 2=blue, 3=orange) of the $|\chi_k\rangle$; from top to bottom: $\mathrm{Re} \chi_a$, $\mathrm{Im} \chi_a$, $\mathrm{Re} \chi_b$, $\mathrm{Im} \chi_b$, $\mathrm{Re} \chi_c$, $\mathrm{Im} \chi_c$.  Same parameters as Figs.~1, 2 and 4.}
\label{fig:xi} 
\end{figure}

\begin{figure}[t!]
\includegraphics[width=0.6\columnwidth]{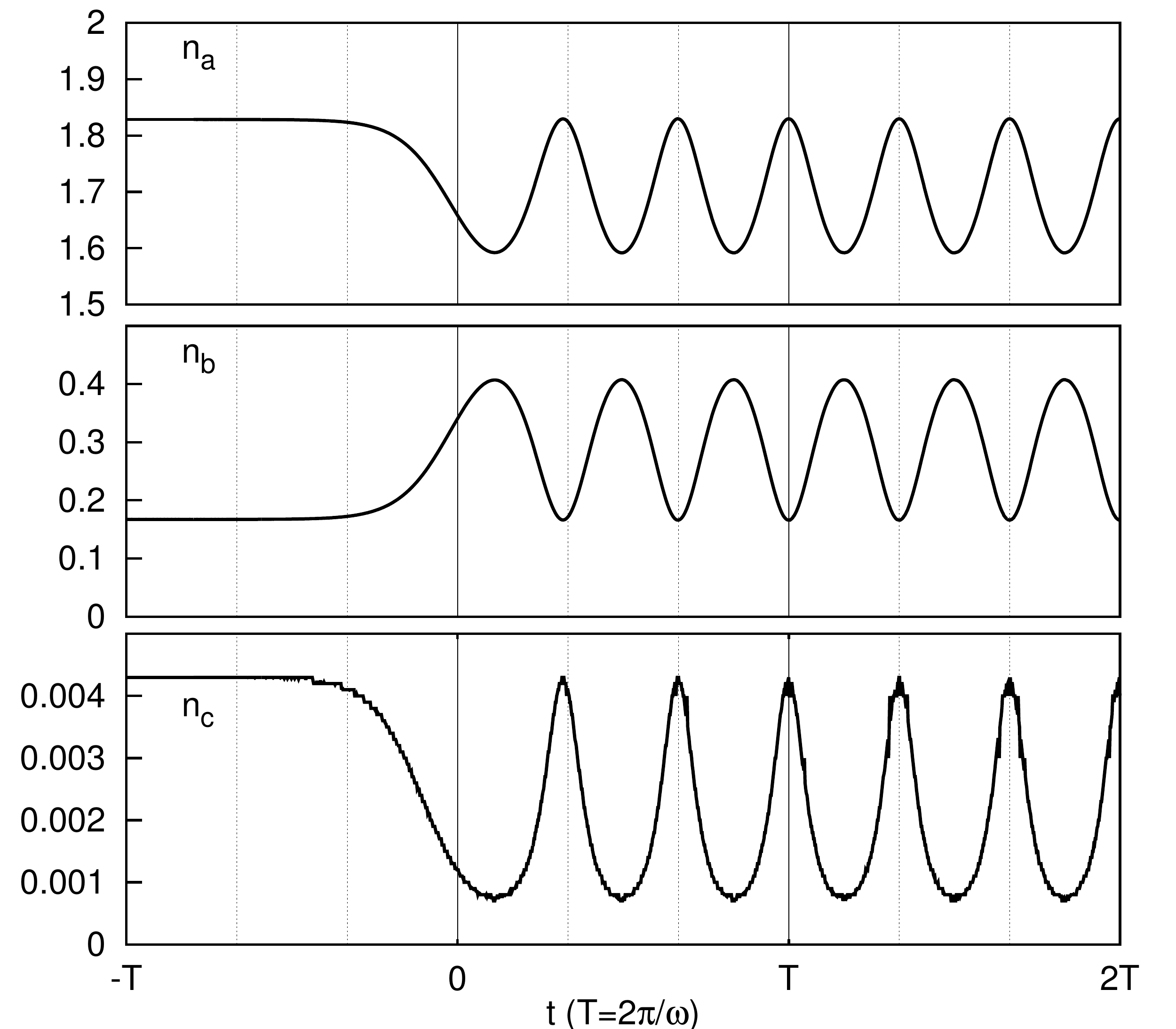}
\caption{Dynamics of the natural orbital occupation numbers $n_a$, $n_b$ and $n_c$ for the same parameters as Figs.~1, 2 and 4.}
\label{fig:nk} 
\end{figure}

The wave function can be expressed in terms of the full set of 10 independent occupation number and angle variables $(A,B|\theta_1, \theta_2, \theta_3, \varphi_1, \varphi_2, \varphi_3| \zeta, \eta, \mu)$ as follows 
\begin{align}
|\Psi\rangle  = \f{1}{\sqrt{2}}\sum_k e^{-i2\zeta_k} \sqrt{n_k} c_{k\uu}^{\dag} c_{k\dd}^{\dag}|0\rangle,
\end{align}
where $\zeta_a = (\mu + \zeta + \eta)/2$, $\zeta_b = (\mu + \zeta - \eta)/2$ and $\zeta_c = (\mu - 2\zeta)/2$.
To obtain the results reported in the Letter, the Schr\"odinger equation was solved in two ways: (i) directly in the complete eigenbasis of many body singlet states, see Eqs.~(\ref{eq:site:basis}), and (ii) via the explicit equations of motion for the 10 occupation number and angle variables. Exactly the same results were obtained in both cases.  The latter equations of motion were derived from the stationary action principle and will be reported elsewhere \cite{s-requist2014}.  Having solved for the 10 occupation number and angle variables, we can construct the dynamics of the states $|\psi_k\rangle = e^{-i\zeta_k} \sqrt{n_k} |\phi_k\rangle$ or, alternatively, the phase-including natural orbitals $|\chi_k\rangle = e^{-i\zeta_k} |\phi_k\rangle$.  The $|\chi_k\rangle$ and $n_k$ are plotted in Figs.~\ref{fig:xi} and \ref{fig:nk} for the same parameters as Figs.~1, 2, and 4.

\begin{figure}[t!]
\includegraphics[width=0.7\columnwidth]{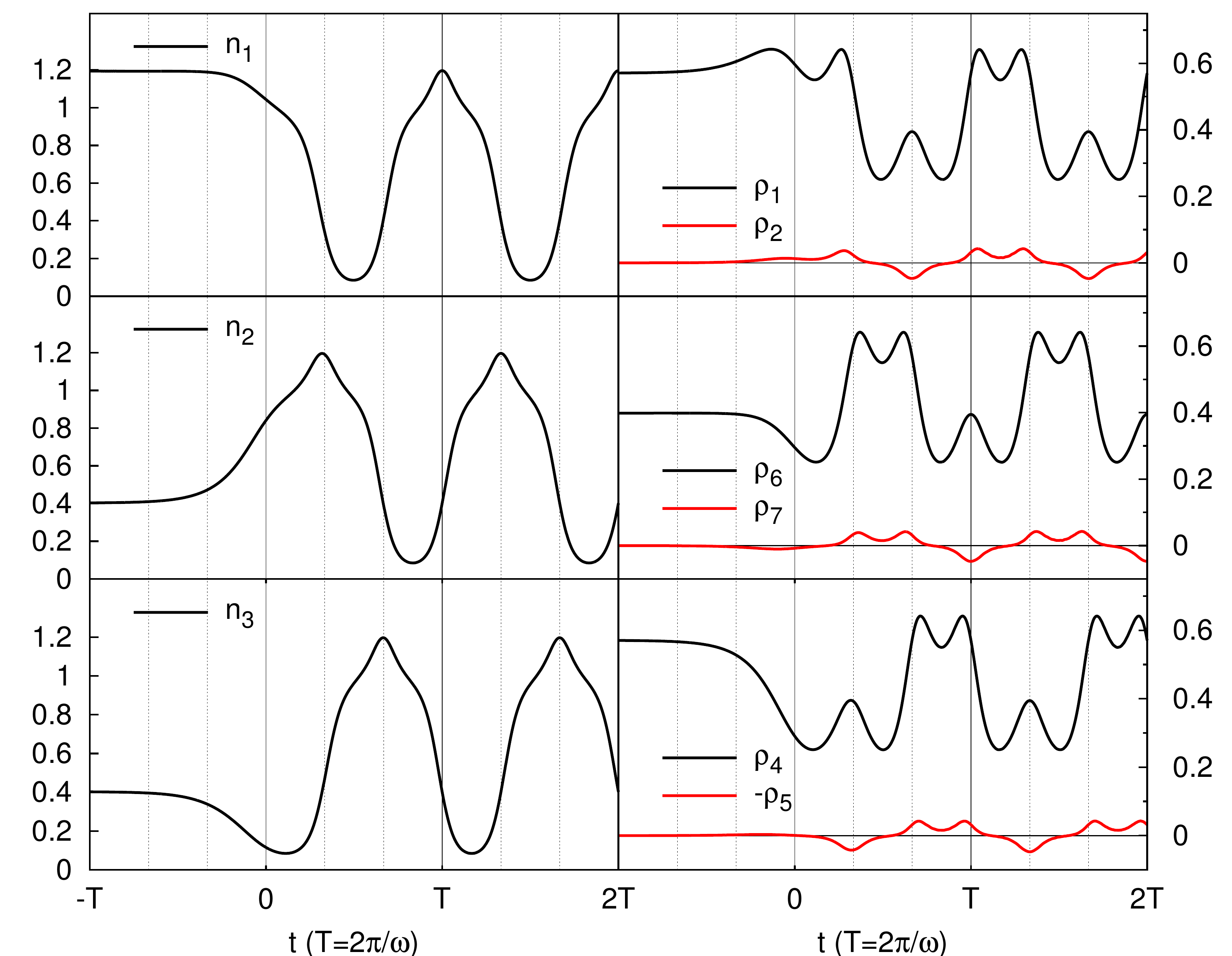}
\caption{Dynamics of the site occupation numbers $n_i$ and off-diagonal elements of $\rho_1$.  Same parameters as Figs.~1, 2 and 4.}
\label{fig:rho1} 
\end{figure}

In analogy with the two-site Hubbard model [cf.~Eq.~(31) of Ref.~\onlinecite{s-requist2010}], the equation of motion for $\rho_1$ can be expressed as a generalized Bloch equation 
\begin{align}
\partial_t \vec{\rho}_1 =  \vec{V} \wedge \vec{\rho}_1 + \vec{U},
\label{eq:bloch}
\end{align}
where the wedge product represents $\f{1}{2} \sum_{ij} C_{ijk} V_i \rho_{1,j}$ and $C_{ijk}$ are the structure constants of the $\mathfrak{su}(3)$ Lie algebra, and $\vec{V}$ and $\vec{U}$ are defined according to $v = \f{1}{2}\vec{V}\cdot \vec{\nu}$ and $u=\f{1}{2}\vec{U}\cdot \vec{\nu}$.  In Fig.~\ref{fig:rho1}, we plot the site occupations and the elements of $\vec{\rho}_1$ corresponding to off-diagonal Gell-Mann matrices, i.e.~$(\rho_1,\rho_2,\rho_4,\rho_5,\rho_6,\rho_7)$, for $U=8$, $\omega_0=1/4$, $\epsilon=4$ and $\phi_0 = 2\pi/3$.  The system reaches a definite Floquet state where all variables are periodic; evidence for the adiabaticity of the ramping wrt the basis of instantaneous Floquet states is given in Sec.~S3.
The function $\vec{\rho}_1(A,B,\theta_1,\theta_2,\theta_3, \varphi_1, \varphi_2, \varphi_3)$ can be inverted analytically, and the reduced geometric phases of the $|\psi_k\rangle$ can be evaluated analytically \cite{s-requist2014}.

Induced gauge geometries and reduced geometric phases are related to the fiber bundles associated with the orbits of $\rho_1$ in case I or $\{|\psi_k\rangle\}$ in case II.  The manifold on which the $\rho_1$ dynamics takes place can be identified with the coadjoint orbit of a Lie group $G$ acting on the dual $\mathfrak{g}^*$ of a Lie algebra $\mathfrak{g}$.  Coadjoint orbits have a natural fiber bundle structure \cite{s-kirillov2004}.  What is interesting about such bundles in the framework of the BBGKY hierarchy is that the fibers are degrees of freedom of higher-order rdms.  For example, a holonomy generated by the dynamics of $\rho_1$ in the base space influences higher-order rdms and hence two-body correlation functions.  Similar arguments apply to $\{|\psi_k\rangle\}$ and analogous sets of variables associated with higher-order rdms \cite{s-requist2012}.  
\bigskip

\noindent {\bf S2. Derivation of an effective many body Hamiltonian with dynamically induced magnetic fields}
\medskip

The induced magnetic gauge potentials studied in the main text appeared in the exact reduced one-body equations of motion, and since those equations have a single-particle form, they can be interpreted as the equations of motion of an effective \textit{noninteracting} system, see Sec.~(S4).  It is interesting, as it gives an alternative perspective, to examine induced gauge potentials at the fully interacting level, i.e.~within a many body approach that retains all interactions.  Floquet theory is one method for doing so, although it appears difficult to obtain analytic results for the present case except within perturbation theory \cite{s-holthaus1992,s-creffield2002a,s-creffield2002b}, e.g.~in the limit $U\ll\omega_0$.  Applying Floquet theory to the present model is an interesting problem for future work.  Here we shall instead consider the adiabatic limit $\omega_0\rightarrow 0$, where we can use standard adiabatic analysis, which has the additional advantage of not being limited to time-periodic dynamics.  We shall find not only generic dynamically-induced gauge potentials but also new types of complex interactions with their own gauge structure.

\begin{figure}[t!]
\includegraphics[width=0.8\columnwidth]{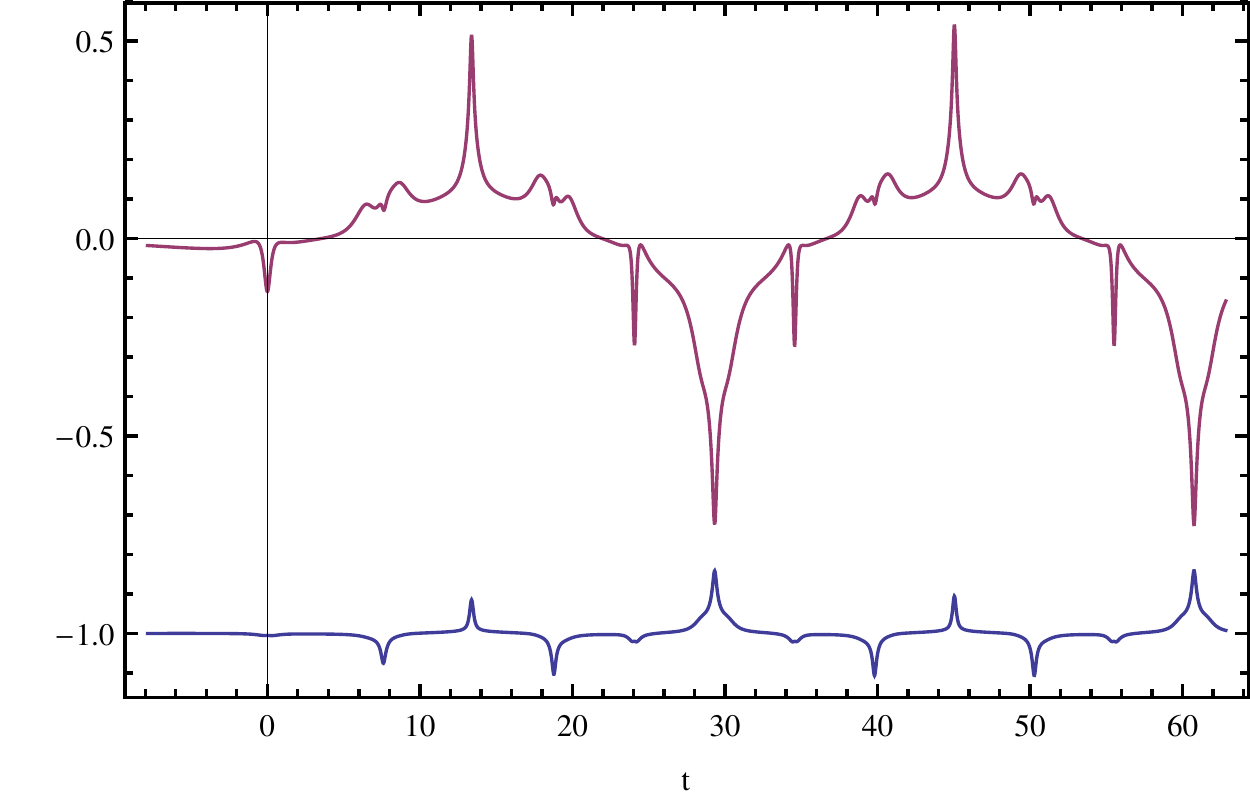}
\caption{Real and imaginary parts of the first-order gauge invariant quantity $t^1_{12} t^1_{23} t^1_{31}$; same parameters as Figs.~1, 2 and 4.}
\label{fig:ttt} 
\end{figure}

In the adiabatic regime, the wave function of a system that starts in the ground state stays close to the instantaneous ground state $|\eta_0(t)\rangle$ throughout the dynamics.  Consider a unitary transformation $U(t)$ to the adiabatic basis $|\eta(t)\rangle = U^{\dag}(t) |\psi(t)\rangle$, where $U(t)$ diagonalizes $H(t)$, i.e.~$U^{\dag} H U = \mathrm{diag}(E_0,E_1\ldots)$.  The dynamical equation for $|\eta(t)\rangle$ is
\begin{align}
i\partial_t |\eta(t)\rangle = H_1(t) |\eta(t)\rangle \label{eq:H1}
\end{align}
with $H_1 = U^{\dag} H U - i U^{\dag} \partial_t U$.  The nonadiabatic term $- i U^{\dag} \partial_t U$ couples the instantaneous eigenstates of $H_1$, thereby inducing a nonvanishing current in the instantaneous ground state of $H_1$ (see Sec.~S5) \cite{s-requist2010}.  The presence of persistent currents in the ground state suggests that the effective Hamiltonian $H_1$ contains induced magnetic gauge potentials.  To prove that it does, it suffices to show that the gauge invariant quantity $t^{1}_{12}t^{1}_{23}t^{1}_{31}=\f{1}{8}(V^1_1-i V^1_2)(V^1_6 -i V^1_7)(V^1_4 -i V^1_5)$ has a nonvanishing argument $\Phi^1 = \mathrm{Arg}\: t^{1}_{12}t^{1}_{23}t^{1}_{31}$.  Here, $t^{1}_{ij}$ are the effective hopping elements of $H_1$ and $V^1_{i} = \mathrm{Tr}(\hat{H}_1\hat{\nu}_i)$ is a vector $\vec{V}^1$, analogous to $\vec{V}$ in Eq.~(\ref{eq:V}).  If $\Phi^1\neq 0$, then $H_1$ contains an induced magnetic flux.  Plots of the real and imaginary parts of $t^{1}_{12}t^{1}_{23}t^{1}_{31}$ in Fig.~\ref{fig:ttt} show this is indeed the case.  The procedure leading to $H_1$ can be iterated to give an $n$th-order adiabatic Hamiltonian $H_n = U_n^{\dag} H_{n-1} U_n - i U_n^{\dag} \partial_t U_n$ \cite{s-berry1987} and further approximations to $\Phi^1$.

We now investigate whether the two-body interactions are also modified by the nonadiabatic coupling.  We will again use Lie algebras, this time $\mathfrak{su}(6)$ instead of $\mathfrak{su}(3)$.  Since the space of two-electron spin singlet states is 6-dimensional, the most general Hamiltonian can be non-redundantly parameterized in terms of the generators of $\mathfrak{su}(6)$.  However, the standard generators of $\mathfrak{su}(6)$, i.e.~the $6\times 6$ matrices with only two nonzero elements, turn out to be linear combinations of one-body and two-body operators.  To see this, first note that e.g.~$\hat{\nu}_2$ can be expressed as
\begin{align}
\hat{\nu}_2 = \sum_{\s} \big(-i c_{1\s}^{\dag} c_{2\s} + i c_{2\s}^{\dag} c_{1\s}\big) = \LB \begin{array}{ccc|ccc} 
0             & 0 &  0 & i\sqrt{2} & -i\sqrt{2} & 0 \\
0             & 0 & -i  &         0    &          0 & 0 \\
0             & i  &  0 &         0    &          0 & 0 \\ \hline
-i\sqrt{2} & 0 &  0 &         0    &          0 & 0 \\
i\sqrt{2}  & 0 &  0 &         0    &          0 & 0 \\
0             & 0 &  0 &         0    &          0 & 0 \end{array} \RB
\end{align}
in the following basis of two-electron states:
\begin{align}
\Lb 1 \Rr &\equiv \Lb \uu \; \dd 0 \Rr \equiv \f{1}{\sqrt{2}} \big( c_{1\uu}^{\dag} c_{2\dd}^{\dag} - c_{1\dd}^{\dag} c_{2\uu}^{\dag} \big) \Lb 0\Rr \nonumber \\
\Lb 2 \Rr &\equiv \Lb \uu 0 \dd \Rr \equiv \f{1}{\sqrt{2}} \big( c_{1\uu}^{\dag} c_{3\dd}^{\dag} - c_{1\dd}^{\dag} c_{3\uu}^{\dag} \big) \Lb 0\Rr \nonumber\\
\Lb 3 \Rr &\equiv \Lb 0 \uu \; \dd \Rr \equiv \f{1}{\sqrt{2}} \big( c_{2\uu}^{\dag} c_{3\dd}^{\dag} - c_{2\dd}^{\dag} c_{3\uu}^{\dag} \big) \Lb 0\Rr \nonumber \\
\Lb 4 \Rr &\equiv \Lb \pa 0 \; 0 \Rr \equiv c_{1\uu}^{\dag} c_{1\dd}^{\dag} \Lb 0\Rr \nonumber\\
\Lb 5 \Rr &\equiv \Lb 0 \pa 0 \Rr \equiv c_{2\uu}^{\dag} c_{2\dd}^{\dag} \Lb 0\Rr \nonumber\\
\Lb 6 \Rr &\equiv \Lb 0 \; 0 \pa \Rr \equiv c_{3\uu}^{\dag} c_{3\dd}^{\dag} \Lb 0\Rr. \label{eq:site:basis}
\end{align}
Since there is no linear transformation among the set of 6 off-diagonal $\hat{\nu}_n$ that makes them coincide with 6 of the standard generators of $\mathfrak{su}(6)$, we conclude that at least some of those generators must correspond to linear combinations of one-body and two-body operators.  Therefore, to build up a complete set of two-body operators that are linearly independent of all one-body operators $\hat{\nu}_n$, we shall have to find  appropriate linear combinations of the standard generators.  By the linear independence of two operators $\hat{A}$ and $\hat{B}$ we mean that $\mathrm{Tr}(\hat{A}\hat{B})=0$, where the trace is taken wrt the basis in Eq.~(\ref{eq:site:basis}), and we normalize all operators such that $\mathrm{Tr}(\hat{A}\hat{A})=10$.

First, consider the on-site Hubbard interactions $U_{iiii} \hat{n}_{i\uu} n_{i\dd}$.They correspond to the diagonal elements $H_{44}$, $H_{55}$, $H_{66}$ of the Hamiltonian in the basis (\ref{eq:site:basis}).  Although they are not orthogonal to $\hat{\nu}_3$, $\hat{\nu}_8$ and $\hat{\nu}_9$ under the trace, we can define the following operators that are:
\begin{align}
\hat{\mu}_1 &= \sqrt{\f{5}{2}} \mathrm{diag}(-1,-1,+1, +1,0,0) \nonumber \\
\hat{\mu}_2 &= \sqrt{\f{5}{2}} \mathrm{diag}(-1,+1,-1, 0,+1,0) \nonumber \\
\hat{\mu}_3 &= \sqrt{\f{5}{2}} \mathrm{diag}(+1,-1,-1, 0,0,+1).
\end{align}

Second, note that the double hopping interactions such as
\begin{align}
W_{1122} \,c_{1\uu}^{\dag} c_{1\dd}^{\dag} c_{2\dd} c_{2\uu} + W_{1122}^*\, c_{2\uu}^{\dag} c_{2\dd}^{\dag} c_{1\dd} c_{1\uu}
\end{align}
are already orthogonal to all one-body operators because they are only nonzero in the lower right block when expressed in the 6-dimensional basis (\ref{eq:site:basis}), i.e.~they only act in the sector of doubly occupied states.  Clearly, the double hopping terms can be put in a one-to-one correspondence with the off-diagonal elements of a set of Gell-Mann matrices for the $\{\Lb 4 \Rr, \Lb 5 \Rr, \Lb 6 \Rr\}$ sector, for example, 
\begin{align*}
\hat{\omega}_1 =  \LB \begin{array}{ccc|ccc} 
0 & 0 & 0 & 0 & 0 & 0 \\
0 & 0 & 0 & 0 & 0 & 0 \\
0 & 0 & 0 & 0 & 0 & 0 \\ \hline
0 & 0 & 0 & 0 & 1 & 0 \\
0 & 0 & 0 & 1 & 0 & 0 \\
0 & 0 & 0 & 0 & 0 & 0 \end{array} \RB \qquad \mathrm{and} \qquad
\hat{\omega}_2 =  \LB \begin{array}{ccc|ccc} 
0 & 0 & 0 & 0 & 0 & 0 \\
0 & 0 & 0 & 0 & 0 & 0 \\
0 & 0 & 0 & 0 & 0 & 0 \\ \hline
0 & 0 & 0 & 0 & -i & 0 \\
0 & 0 & 0 & i & 0 & 0 \\
0 & 0 & 0 & 0 & 0 & 0 \end{array} \RB.
\end{align*}
The double hopping amplitudes define a nontrivial gauge invariant loop quantity $W_{1122} W_{2233} W_{3311}$ like the hopping amplitudes.  In the two-site Hubbard model, the complex phase of the double hopping amplitude $W_{1122}$ was found to couple strongly to the dynamics of the occupation numbers $n_k$ and relative phases $\zeta_k$ \cite{s-requist2011}.

\begin{figure}[t!]
\includegraphics[width=0.52\columnwidth]{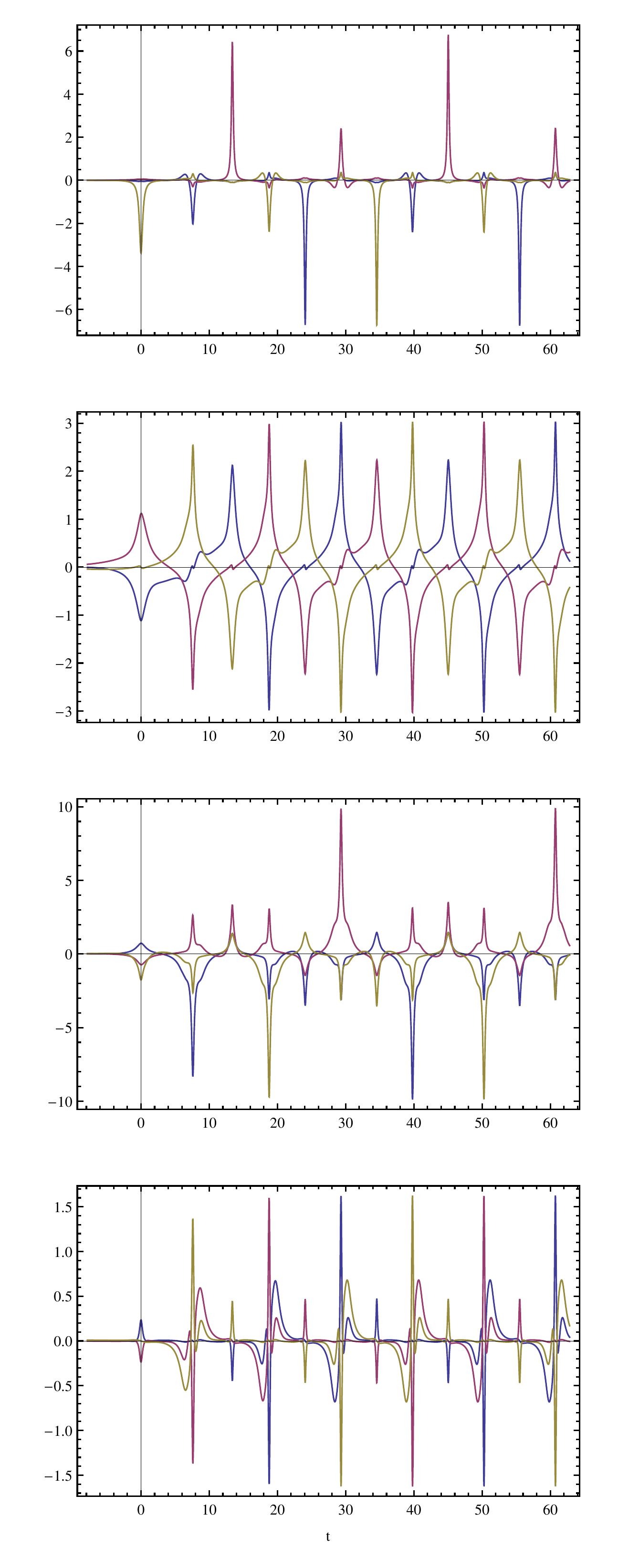}
\caption{Dynamically induced double hopping (top panel) and correlated hopping (lower three panels) amplitudes corresponding to the operators $\hat{\omega}_n$, $\hat{\sigma}_n$, $\hat{\tau}_n$, and $\hat{\chi}_n$, respectively, obtained from the first-order adiabatic Hamiltonian.  In each case, only the amplitudes corresponding to imaginary generating functions are nonzero.  Same parameters as Figs.~1, 2 and 4.}
\label{fig:WSTX} 
\end{figure}

Third, consider correlated hopping terms such as
\begin{align}
|1\rangle \langle 4| + |4\rangle \langle 1| &= S_{1211} \f{1}{\sqrt{2}} (c_{1\uu}^{\dag} c_{2\dd}^{\dag} - c_{1\dd}^{\dag} c_{2\uu}^{\dag}) c_{1\dd} c_{1\uu} + H.c. 
\end{align}
We can decompose these terms into three types of interactions.  The first, which we  denote by $\hat{\sigma}_n$, are nonzero only in the upper right and lower left blocks, e.g.
\begin{align*}
\hat{\sigma}_1 = \LB \begin{array}{ccc|ccc} 
0             & 0 &  0 & \sqrt{\f{5}{2}} & -\sqrt{\f{5}{2}} & 0 \\
0             & 0 &  0 &            0 &             0 & 0 \\
0             & 0 &  0 &            0 &             0 & 0 \\ \hline
\sqrt{\f{5}{2}}  & 0 &  0 &            0 &             0 & 0 \\
-\sqrt{\f{5}{2}} & 0 &  0 &            0 &             0 & 0 \\
0             & 0 &  0 &            0 &             0 & 0 \end{array} \RB, \qquad
\hat{\sigma}_2 = \LB \begin{array}{ccc|ccc} 
0             & 0 &  0 & i\sqrt{\f{5}{2}} & i\sqrt{\f{5}{2}} & 0 \\
0             & 0 &  0 &            0 &             0 & 0 \\
0             & 0 &  0 &            0 &             0 & 0 \\ \hline
-i\sqrt{\f{5}{2}}  & 0 &  0 &            0 &             0 & 0 \\
-i\sqrt{\f{5}{2}} & 0 &  0 &            0 &             0 & 0 \\
0             & 0 &  0 &            0 &             0 & 0 \end{array} \RB,
\end{align*}
or, as an operator in second quantization,  
\begin{align}
\hat{\sigma}_1 = \sqrt{\f{5}{2}} (n_{1\uu} - n_{2\uu}) (c_{2\dd}^{\dag} c_{1\dd} + c_{1\dd}^{\dag} c_{2\dd}) + \sqrt{\f{5}{2}} (n_{1\dd} - n_{2\dd}) (c_{2\uu}^{\dag} c_{1\uu} + c_{1\uu}^{\dag} c_{2\uu}).
\end{align}
The second type of interaction $\hat{\tau}_n$ has a form such as 
\begin{align*}
\hat{\tau}_1 = \LB \begin{array}{ccc|ccc} 
0                        & 0 &  0 & -\f{1}{\sqrt{2}} &  -\f{1}{\sqrt{2}} & 0 \\
0                        & 0 &  2 &                        0 &                         0 & 0 \\
0                        & 2 &  0 &                        0 &                         0 & 0 \\ \hline
-\f{1}{\sqrt{2}} & 0 &  0 &                        0 &                         0 & 0 \\
-\f{1}{\sqrt{2}} & 0 &  0 &                        0 &                         0 & 0 \\
0                        & 0 &  0 &                        0 &                         0 & 0 \end{array} \RB, \qquad
\hat{\tau}_2 = \LB \begin{array}{ccc|ccc} 
0                        & 0 &  0 & -i\f{1}{\sqrt{2}} &  i\f{1}{\sqrt{2}} & 0 \\
0                        & 0 &  -i2 &                        0 &                         0 & 0 \\
0                        & i2 &  0 &                        0 &                         0 & 0 \\ \hline
-\f{1}{\sqrt{2}} & 0 &  0 &                        0 &                         0 & 0 \\
-\f{1}{\sqrt{2}} & 0 &  0 &                        0 &                         0 & 0 \\
0                        & 0 &  0 &                        0 &                         0 & 0 \end{array} \RB,
\end{align*} 
or
\begin{align}
\hat{\tau}_1 = \f{1}{2} (- n_{1\uu} - n_{2\uu} + n_{3\uu}) (c_{2\dd}^{\dag} c_{1\dd} + c_{1\dd}^{\dag} c_{2\dd}) + \f{1}{2} (- n_{1\dd} - n_{2\dd} + n_{3\dd}) (c_{2\uu}^{\dag} c_{1\uu} + c_{1\uu}^{\dag} c_{2\uu}).
\end{align}
The third type of interaction $\hat{\chi}_n$ corresponds to the elements $H_{i,7-i}$ $(i=1\ldots 6)$ and leads to terms such as 
\begin{align}
n_{3\uu} (c_{2\dd}^{\dag} c_{1\dd} + c_{1\dd}^{\dag} c_{2\dd}) + n_{3\dd} (c_{2\uu}^{\dag} c_{1\uu} + c_{1\uu}^{\dag} c_{2\uu}),
\end{align}
All of the correlated hopping operators $\hat{\sigma}_n$, $\hat{\tau}_n$ and $\hat{\chi}_n$ are orthogonal to $\hat{\nu}_1$ under the trace.  Numerical calculations confirm that all of the dynamically induced interaction operators with imaginary matrix elements (e.g.~$\hat{\omega}_2$, $\hat{\sigma}_2$ and $\hat{\tau}_2$) are generically present in $H_1$.  Figure \ref{fig:WSTX} shows the amplitudes of double hopping terms $\hat{\omega}_n$ and correlated hopping $\hat{\sigma}_n$, $\hat{\tau}_n$ and $\hat{\chi}_n$ obtained from the Hamiltonian $H_1$.  Correlated hopping terms similar to these were studied for solids with intermediate valency \cite{s-foglio1979} and, recently, for ultracold atoms in optical lattices \cite{s-dutta2011,s-rapp2012,s-dilibreto2014}.

The operators $(\hat{\nu}_n|\hat{\mu}_n, \hat{\s}_n, \hat{\tau}_n, \hat{\omega}_n, \hat{\chi}_n)$ form a complete set of generators for the $\mathfrak{su}(6)$ algebra.  As expected, there are a total of $36=6^2$ independent operators.  One advantage of these operators is that they can be used to explicitly separate one-body and two-body degrees of freedom.  The most general two-body rdm $\hat{\rho}_2$ can be expanded uniquely in terms of the one-body and two-body operators.  The above approach to orthogonalizing one-body and two-body operators can be extended to more complex systems and might be useful in the study of general lattice models.  It might also be useful in identifying appropriate gauge invariant quantities for density functional-type theories, e.g.~current density functional theory \cite{s-vignale1987}, where the basic independent variables should be gauge invariant.  There are likely connections with the geometry of entanglement (see Ref.~\onlinecite{s-bengtsson2006} and references therein) in quantum information theory.
\bigskip 

\noindent {\bf S3. Verifying adiabatic ramping and tracking the adiabatically continued quasienergies}
\medskip

Starting from a given stationary state, perhaps the simplest way to bring a system to a Floquet state is to turn on the periodic driving slowly enough that the system has a chance to adjust and adiabatically build up periodically oscillating components.  A convenient way to formulate this mathematically is to send the initial time back to $-\infty$ and employ an adiabatic ramping function such as $f(t)=(1/2)(1+\tanh\alpha t)$, which turns the perturbation on over a slow time scale $\tau=\alpha^{-1}$ and approaches a constant value of 1 as $t\rightarrow\infty$.  In this way the Hamiltonian, although not perfectly periodic during the ramping, approaches a periodic function in the limit $t\rightarrow\infty$.

If the ramping is successful, the system will have evolved adiabatically from a given stationary state to a given Floquet state.  By varying the details of the time-periodic Hamiltonian and the ramping function, one can map out the set of Floquet states that are reachable from the initial stationary state.  A version of the adiabatic theorem has been proved for adiabatically varied time-periodic Hamiltonians \cite{s-young1970}.  The key point is that the adiabatic eigenenergies which enter in the adiabatic theorem get replaced by the instantaneous quasienergies, which are the quasienergies one would obtain by solving Eq.~(\ref{eq:KOmega}) below for the time-periodic Hamiltonian with a ``frozen'' value of the ramping function.  One can then adiabatically continue these instantaneous quasienergies by varying the parameters of the ramping.  In doing so, the Floquet state corresponding to a given quasienergy is adiabatically transported in the space of Floquet states.  Any two states that can be connected in this way will be called \textit{adiabatically connnected}.

\begin{figure}[t!]
\includegraphics[width=0.65\columnwidth]{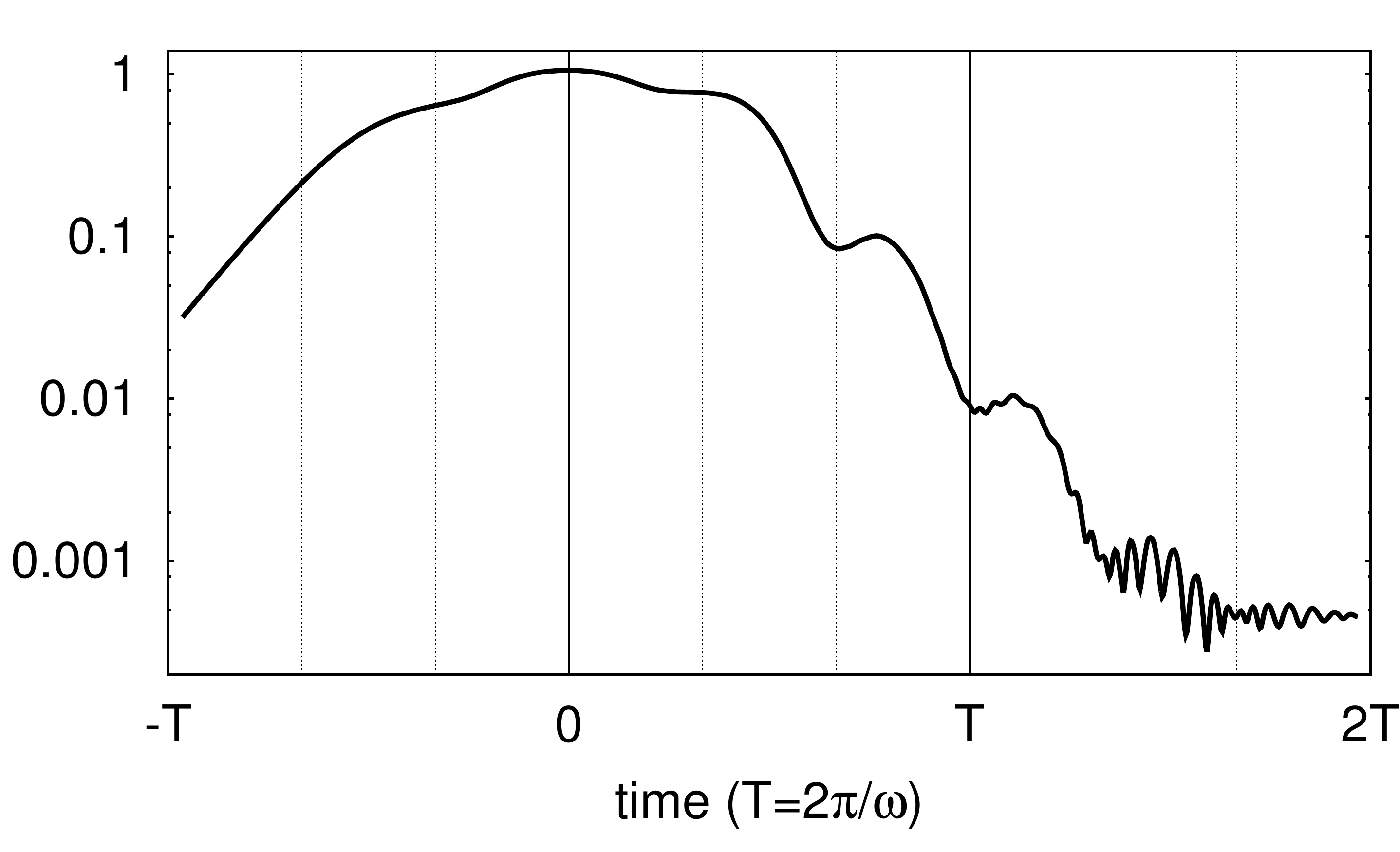}
\caption{Error $\delta(t) = ||\xi(t+T)-\xi(t)||$ in the Floquet state $|\Psi(t)\rangle = e^{-i\int_{t_0}^t\Omega(s)ds} |\xi(t)\rangle$ for the parameters in Figs.~1, 2 and 4.  In the limit $t\rightarrow \infty$, the error approaches 0.00045.}
\label{fig:delta} 
\end{figure}

\begin{figure}[t!]
\includegraphics[width=0.7\columnwidth]{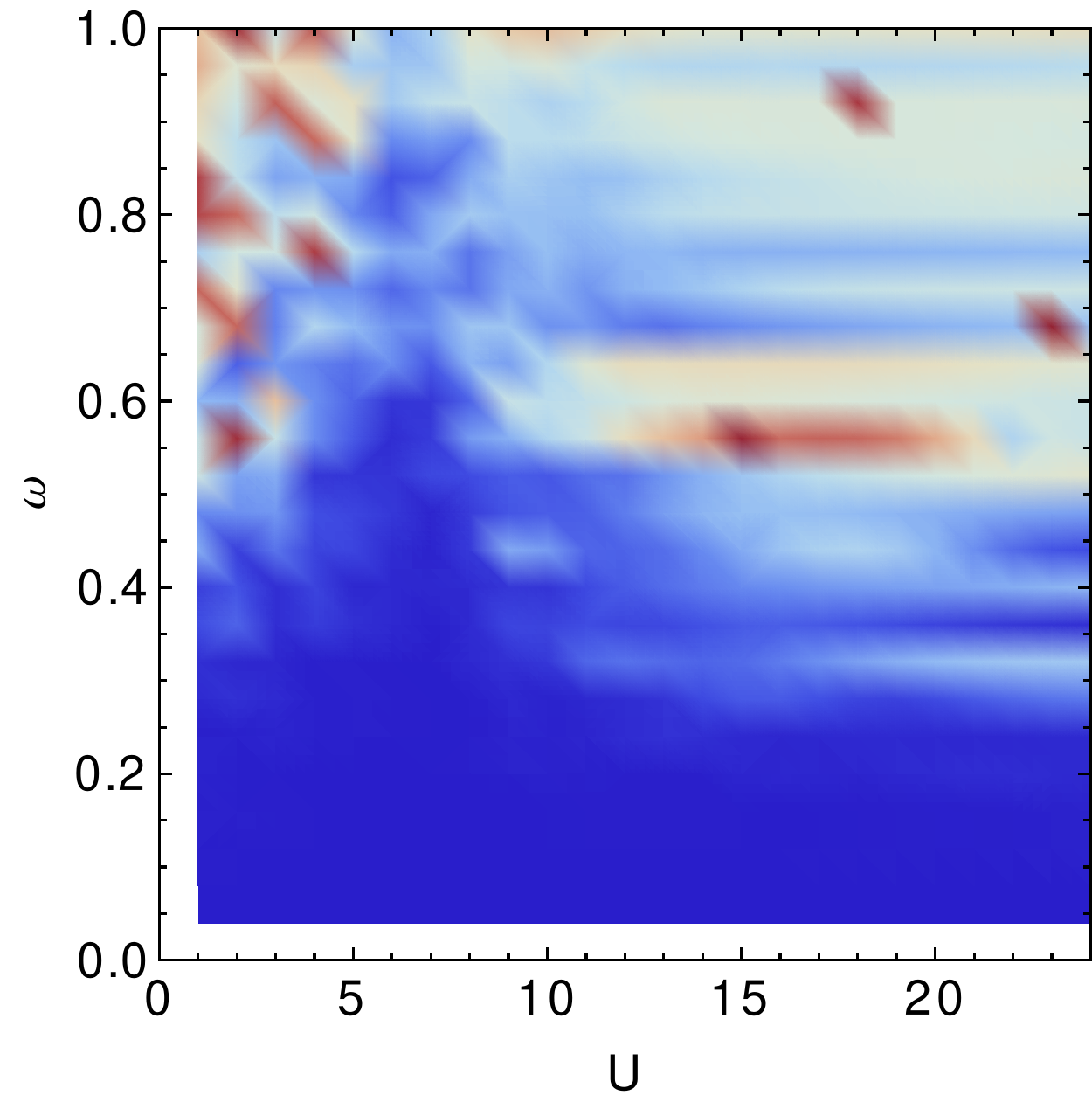}
\caption{Error $\delta = \lim_{t\rightarrow\infty} ||\xi(t+T)-\xi(t)||$ as a function of $(U,\omega_0)$ for $\epsilon=4$ and $\phi_0=0$. Scale from 0-2 (blue-red).}
\label{fig:delta:uomega} 
\end{figure}

Two states might not be adiabatically connected if the quasienergy in question undergoes any avoided crossings with other quasienergies during the ramping.  In analogy with Landau-Majorana-Zener transitions between adiabatic eigenstates, there may be appreciable nonadiabatic transitions at such avoided crossings (see e.g.~\cite{s-drese1999}).  Here we demonstrate numerically that our system with the ramping given by Eqs.~(\ref{eq:driving:s}) does indeed evolve adiabatically to a Floquet state to high accuracy.  The figure of merit is the periodicity of the factor $|\xi(t)\rangle$.  Deviations from periodicity are measured by $\delta = \limsup_{t\rightarrow\infty} ||\xi(t+T)-\xi(t)||$.  The quantity $||\xi(t+T)-\xi(t)||$ is plotted in Fig.~\ref{fig:delta} for the same parameters that were used in Figs.~1, 2 and 4, namely $\alpha=0.11$, $\epsilon=4$, $\omega_0=0.2$, $U=7$ and $\phi_0=0$, and in the limit $t\rightarrow\infty$ it approaches $\delta \approx .00045$.  The error in the quasienergy is $\mathcal{O}(\delta^2)$.  To convey a sense of the effectiveness of adiabatic ramping globally in parameter space, in Fig.~\ref{fig:delta:uomega} we plot $\delta$ as a function of $(U,\omega_0)$ for $\epsilon=4$, $\alpha = \mathrm{min}(\omega_0,1/8)$, $\phi_0=0$ and $t_0=-3T/2$.

\begin{figure}[t!]
\includegraphics[width=0.9\columnwidth]{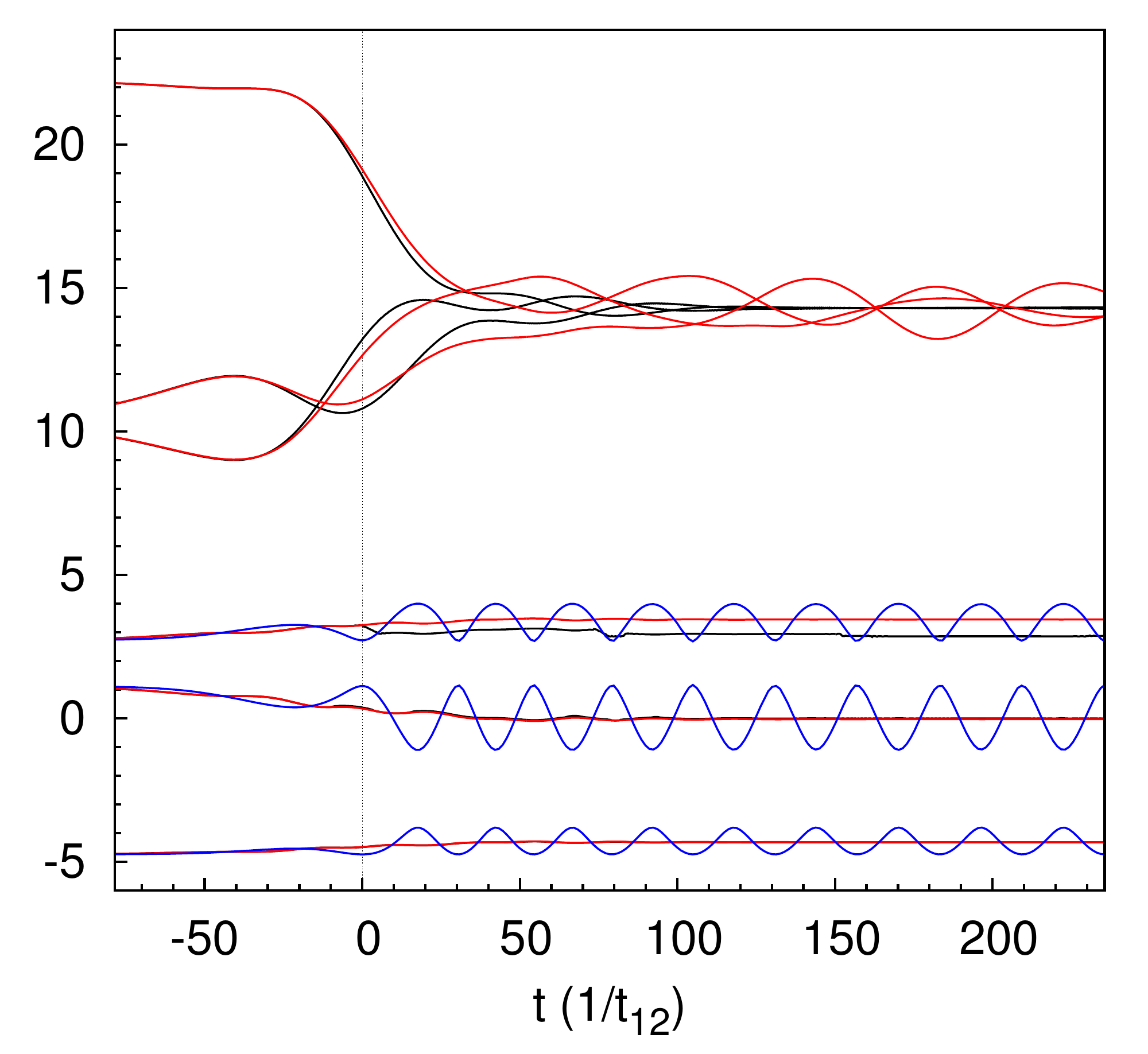}
\caption{Time dependence of the six quasienergies $\Omega_{nT}(t)$ (black) during the ramping is shown for $\alpha=0.03$, $\omega_0=0.08$, $\epsilon=4$ and $U=14$.  The adiabatic eigenenergies $E_n(t)$ (blue) and their running time averages $E_{nT}(t)$ (red) are shown for comparison.}
\label{fig:quasienergies} 
\end{figure}

Time dependent quasienergies $\Omega_n(t)$ can be defined by propogating from the $n$th stationary state of the initial Hamiltonian $H(-\infty)$.  In order to investigate whether there are avoided crossings of the quasienergies during adiabatic ramping, the running time averages $\Omega_{nT}(t)$ are plotted in Fig.~\ref{fig:quasienergies} in the adiabatic regime.  Also shown are the adiabatic eigenenergies $E_n(t)$ and their running time averages $E_{nT}(t)$.  All running time averages approach constants in the limit $t\rightarrow\infty$.  There is apparently a strong level attraction between the quasienergies of the highest sector of states, but the quasienergies of the Floquet states obtained from the lowest three states remain separated from each other by an energy gap uniformly throughout the ramping.  

\newpage

\noindent {\bf S4. Modified continuity equation for an effective noninteracting ensemble}  
\medskip

The set of natural orbitals $\{|\phi_k\rangle\}$ and their occupation numbers $\{n_k\}$ can be interpreted as defining a noninteracting ensemble system \cite{s-requist2008,s-requist2010}.  The natural orbitals have been augmented by phase factors $e^{-i\zeta_k}$, which has the advantage that it allows all elements of the effective single-particle Hamiltonian to be uniquely defined and it incorporates into the ensemble system phase variables $\zeta_k$ that are important for the time dependence of the occupation numbers \cite{s-giesbertz2010}.  There is a geometric motivation for further augmenting the states with an amplitude factor $\sqrt{n_k}$, giving the single-particle states $|\psi_k\rangle = e^{-i\zeta_k} \sqrt{n_k} |\phi_k\rangle$ \cite{s-requist2012}.  Propagating the states $|\psi_k\rangle$, as we have done here, is clearly equivalent to simultaneously propagating the equations of motion for the $n_k$ and the effective Schr\"odinger equation defined in \cite{s-giesbertz2010}.  However, there is an important difference that one should be aware of.  Since the modulus of $|\psi_k\rangle$ is time dependent, the single-particle Hamiltonian must be non-Hermitian, and therefore the continuity equation is modified.

For a unitary noninteracting system on a lattice, the continuity equation is
\begin{align}
\partial_t n_{\mu} = -\sum_{\nu} J_{\mu\nu}, \label{eq:continuity}
\end{align}
where $J_{\mu\nu} = \mathrm{Tr}(\hat{J}_{\mu\nu} \hat{\rho}_1) = 2\,\mathrm{Im}\, t_{\mu\nu} \rho_{1,\nu\mu}$ and the current operator is $\hat{J}_{\mu\nu}=-i t_{\mu\nu} c_{\mu}^{\dag} c_{\nu} + i t_{\nu\mu} c_{\nu}^{\dag} c_{\mu}$.  For a noninteracting ensemble with non-Hermitian Hamiltonian, the dynamics is nonunitary and the continuity equation becomes
\begin{align}
\partial_t n_{\mu} 
&= \sum_k \f{1}{i} \langle \psi_k | \hat{n}_{\mu} \hat{h} | \psi_k \rangle - \f{1}{i} \langle  \psi_k | \hat{h}^{\dag} \hat{n}_{\mu} | \psi_k \rangle \nonumber \\
&= \sum_k \f{1}{i} \langle \psi_k | [\hat{n}_{\mu}, \hat{h}^{\chi}] | \psi_k \rangle + \langle  \psi_k | \{\hat{n}_{\mu}, \hat{h}^{\xi}\}  | \psi_k \rangle \nonumber \\
&= -\sum_{\nu} J_{\mu\nu} + \mathrm{Tr}(\{\hat{n}_{\mu}, \hat{h}^{\xi}\} ), \label{eq:continuity:2}
\end{align}
where $h=h^{\chi} + i h^{\xi}$ is divided into Hermitian and skew-Hermitian terms ($h^{\chi}$ and $h^{\xi}$ are Hermitian).  The second term in Eq.~(\ref{eq:continuity:2}) is a correction due to the non-Hermiticity of $h$, which acts as an additional source/drain. 
%
%

Despite this modification of the continuity equation, the noninteracting system exactly reproduces the current and all one-body observables of the interacting many body system, since $\rho_1 = \sum_k |\psi_k\rangle \langle \psi_k| = \sum_k n_k |\phi_k\rangle \langle \phi_k|$.  Plots of the instantaneous circulating current $J(t)T = \f{T}{3} (J_{12}(t) +J_{23}(t) + J_{31}(t)) = \f{2T}{3} t_{12} (\rho_2(t)+\rho_7(t)-\rho_5(t))$ together with its running time average, the pumped charge $Q(t)=J(t)T$, are shown in Fig.~\ref{fig:Q}.

\bigskip

\noindent {\bf S5. Derivation of Eq.~(13) and its relationship to the stationary principle for the quasienergy} 
\medskip

The following derivation of Eq.~(13) is essentially equivalent to the derivation of a similar formula for charge pumping in superconducting circuits \cite{s-russomanno2011} but the phase has a different physical meaning.  Start from the eigenvalue equation  
\begin{align}
K|\xi \rangle = \Omega |\xi\rangle, \qquad K=H-i\partial_t,\label{eq:KOmega}
\end{align}
which determines the periodic factor of a given Floquet state $|\Psi\rangle = e^{-i\Omega t}|\xi\rangle$.  Taking the partial derivative with respect to $\Phi$, using the definition $\hat{J} = -e\, \partial H/\partial \Phi$, and multiplying by $\langle \xi|$ gives
\begin{align}
-\f{1}{e} J + \Big< \xi \Big| K \Big|\f{\partial \xi}{\partial \Phi} \Big> = \f{\partial \Omega}{\partial \Phi} + \Omega \Big< \xi \Big| \f{\partial \xi}{\partial \Phi} \Big>.
\end{align}
The second and fourth terms are seen to cancel after averaging over one period $T$ and integrating by parts, which gives the desired result
\begin{align}
\f{Q}{e} = \f{J_T T}{e} = - \f{\partial \Omega T}{\partial \Phi}. \label{eq:13}
\end{align}
It is clear that $(\Phi,J)$ are conjugate variables and that the above derivation can be generalized to any such pair of conjugate variables $(Q,P)$, e.g.~$(\epsilon_i,n_i)$ or $(V_i,\rho_{1,i})$.  

Equation (13) is closely related to a stationary principle for the quasienergy function $\Omega(J_T)$, which is a special case of the stationary principle for the quasienergy functional $\Omega[\rho_{1T}]$.   A stationary principle for $\Omega[\rho_{1T}]$ to second order in a harmonic perturbation was proved in Ref.~\onlinecite{s-requist2009}, which focused on the general many electron problem in cases where the spectrum has a continuum component.  The problem of defining such a stationary principle simplifies when the Hilbert space is finite dimensional, as in the present case, and the arguments of Ref.~\onlinecite{s-requist2009} can be extended to define a stationary principle to all orders.  Note that the three-fold symmetry of the Floquet state in the present model greatly reduces the number of independent parameters of $\rho_{1T}$.  We now establish a relationship between Eq.~(13) and the stationary principle for $\Omega(J_T)$.

We begin by defining the quasienergy as a function of a constant externally applied flux $\Phi$ according to 
\begin{align}
\Omega(\Phi) = \langle \xi(\Phi,t) | H(\Phi,t)-i\partial_t | \xi(\Phi,t) \rangle,
\end{align}
where $|\xi(\Phi,t)\rangle$ is the $T$-periodic factor of the steady Floquet state $|\Psi\rangle = e^{-i\Omega t} |\xi\rangle$.  The flux is added to the Hamiltonian in Eq.~(5) by making the hopping amplitudes complex.  If $\Omega(\Phi)$ is convex on a given interval $\mathcal{D}$, then we can define the Legendre transform
\begin{align}
\mathcal{F}(J_T) = \min_{\Phi\in\mathcal{D}} \Phi J_T + \Omega(\Phi).
\end{align}
Unlike the internal energy functional $F[n]$ in density functional theory, $\mathcal{F}(J_T)$ is not a universal functional of $J_T$ since it depends on the details of the time periodic driving as well as which Floquet state the system is in.  Convexity implies a 1:1 relationship $\Phi \leftrightarrow J_T$ on $\mathcal{D}$, so we can write $\mathcal{F}(J_T) = \Phi J_T + \Omega(\Phi)$, substituting $\Phi=\Phi(J_T)$. 

For fixed $\Phi\in\mathcal{D}$, we define the quasienergy $\Omega_{\Phi}(J) = -\Phi J_T + \mathcal{F}(J_T)$ which satisfies a minimum principle because $\mathcal{F}(J_T)$ is convex.  At the minimum, we have
\begin{align}
\f{\partial \mathcal{F}}{\partial J_T} = \Phi. \label{eq:F:1}
\end{align}
Using the chain rule gives
\begin{align}
\f{\partial \mathcal{F}}{\partial J_T} = \f{\partial \mathcal{F}}{\partial \Phi} \f{\partial \Phi}{\partial J_T} + J_T \f{\partial \Phi}{\partial J_T} + \Phi.  \label{eq:F:2}
\end{align}
Therefore, equations~(\ref{eq:F:1}) and (\ref{eq:F:2}) together imply Eq.~(\ref{eq:13}).

The definition of the Legendre transform $\mathcal{F}(J_T)$ is valid over any interval of $\Phi$ on which $\Omega(\Phi)$ is convex.  If $\Omega(\Phi)$ is concave, then one defines the Legendre transform analogously using $-\Omega(\Phi)$.  Figure~\ref{fig:convexity} shows that $\Omega(\Phi)$ is convex on an interval $(-0.8,\f{\pi}{2})$ for $U=2$ and $U=3$ and with all other parameters the same as in Figs.~1, 2 and 4.  As $U$ increases through a critical value $U_c \gtrsim 4$, $\Omega(\Phi)$ switches from convex to concave.  This implies it is not possible to define a single Laplace transform that is valid globally in parameter space.  It is also likely that $\Omega(\Phi)$ is not uniformly convex (or concave) over the full range $(-\pi,\pi)$, but we have not been able to verify this since the efficiency of adiabatic ramping to a Floquet state degrades dramatically when $|\Phi|>\f{\pi}{2}$.  In multivariate cases, $\Omega$ might be neither convex nor concave in some parameter regimes, e.g.~this is expected when $\omega_0$ is greater than the first excitation energy \cite{s-langhoff1972}.  

The first-order adiabatic quasienergy $\Omega^1$ can be derived from the instantaneous ground state $|\psi_{gs}^{1}\rangle$ of the Hamiltonian $H_1$ in Sec.~S2.  In this case, one can verify analytically that $\partial \Omega^1_{geo}/\partial \Phi = - J_T$, where $\Omega^1_{geo} = -i \f{1}{T} \int_t^{t+T} \langle \xi|\partial_s \xi\rangle ds$.  Figure \ref{fig:kinematic} shows how $\Omega_{geo}$ depends on $\omega_0$ and the Hubbard interaction $U$.

\begin{figure}[t!]
\includegraphics[width=\columnwidth]{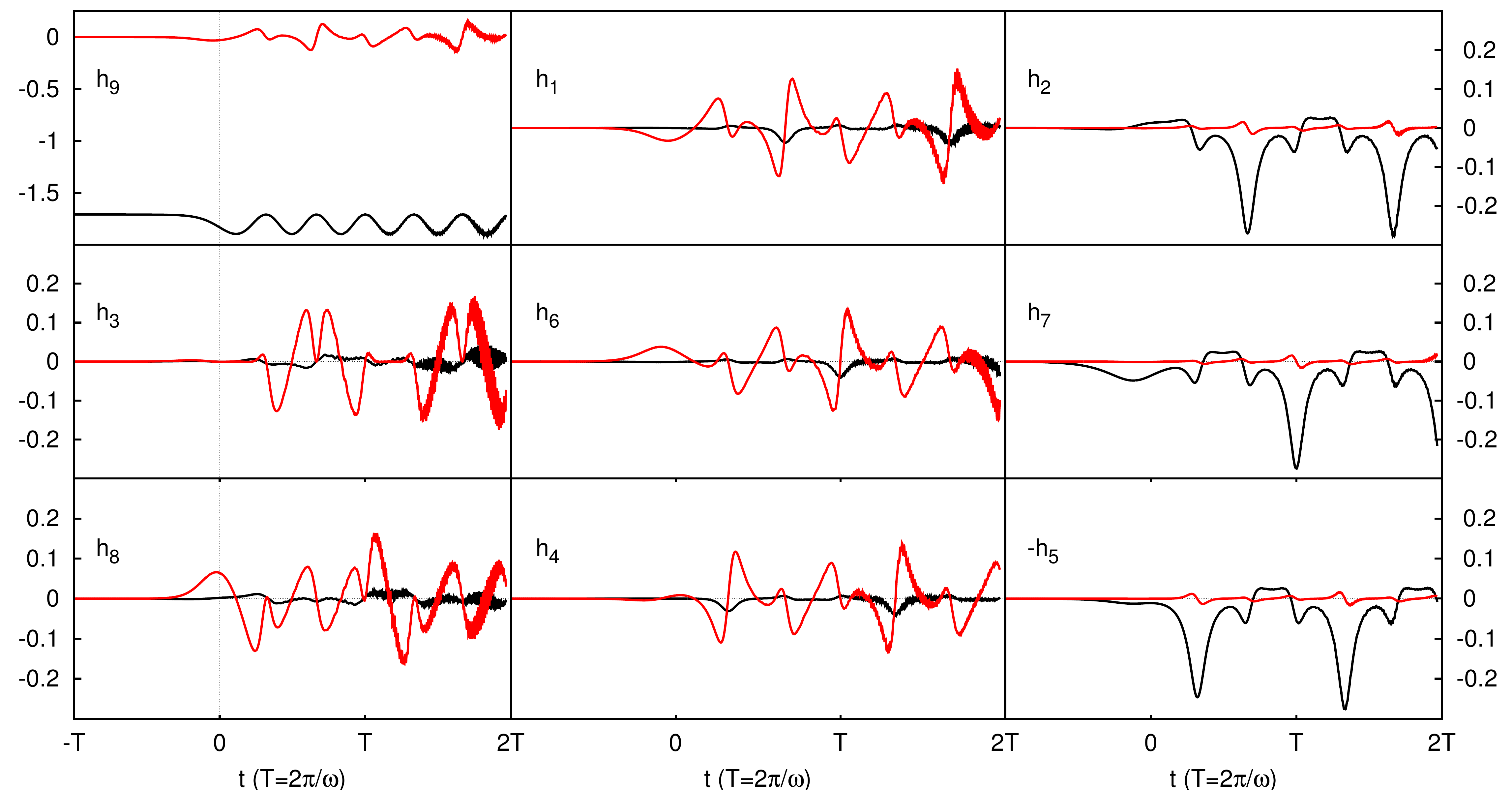}
\caption{Time dependence of the Hamiltonian $h$ for the same parameters as Figs.~1, 2 and 4.}
\label{fig:hpsi} 
\end{figure} 

\begin{figure}[t!]
\includegraphics[width=\columnwidth]{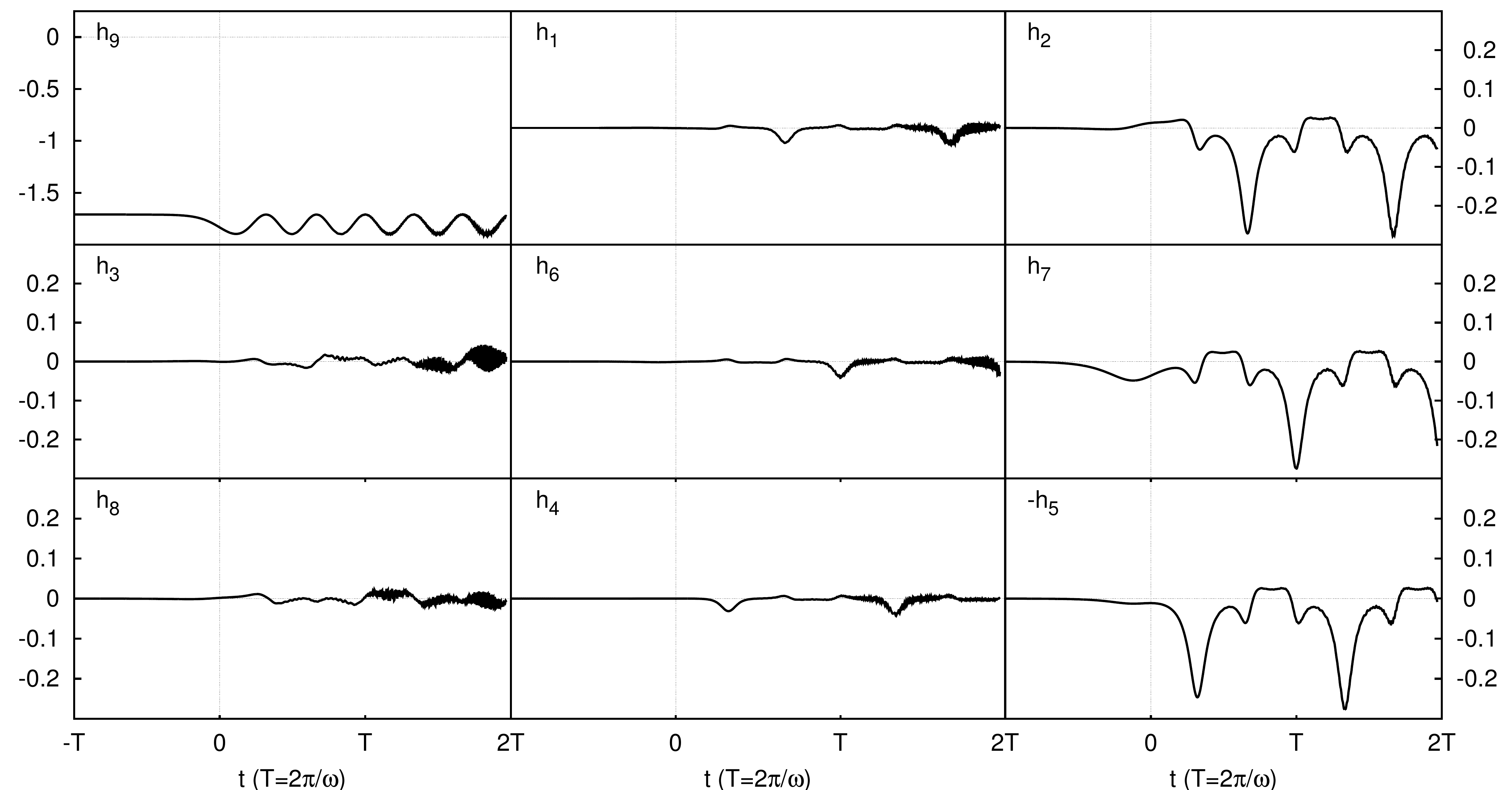}
\caption{Time dependence of the Hamiltonian $h^{\chi}$ for the same parameters as Figs.~1, 2 and 4.}
\label{fig:hchi} 
\end{figure}

\begin{figure}[t!]
\includegraphics[width=0.8\columnwidth]{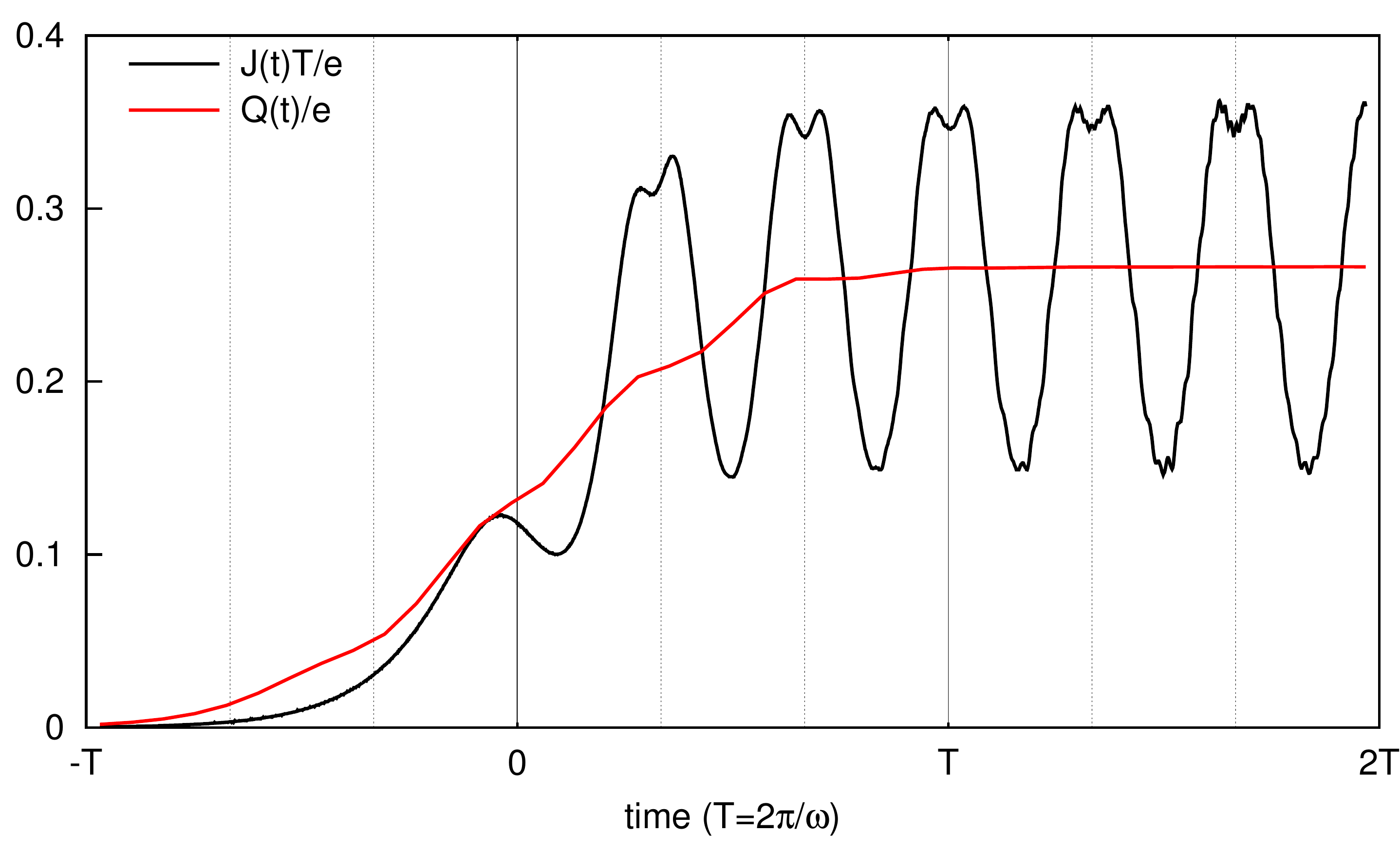}
\caption{Dynamics of the instantaneous circulating current $J(t)T/e$ and its running time average $Q(t)$.}
\label{fig:Q} 
\end{figure} 

\begin{figure}[t!]
\includegraphics[width=0.7\columnwidth]{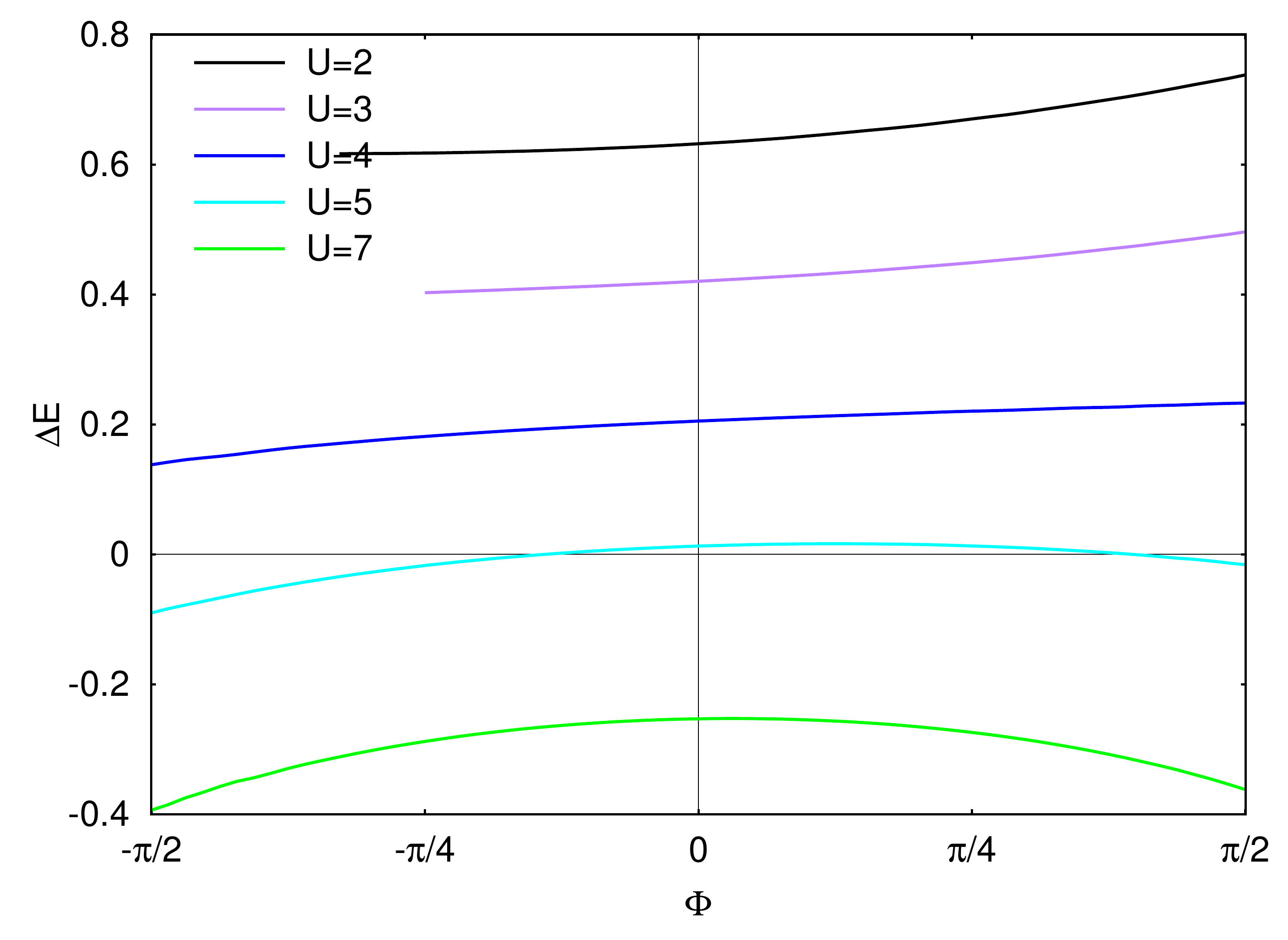}
\caption{Convexity of the quasienergy $\Omega(\Phi)$ with respect to an externally applied magnetic flux $\Phi$ at various values of $U$.  Parameters are $\alpha=0.11$, $\epsilon=4$, $\omega_0=0.2$, $\phi_0=0$.}
\label{fig:convexity} 
\end{figure} 

\begin{figure}[t!]
\includegraphics[width=0.8\columnwidth]{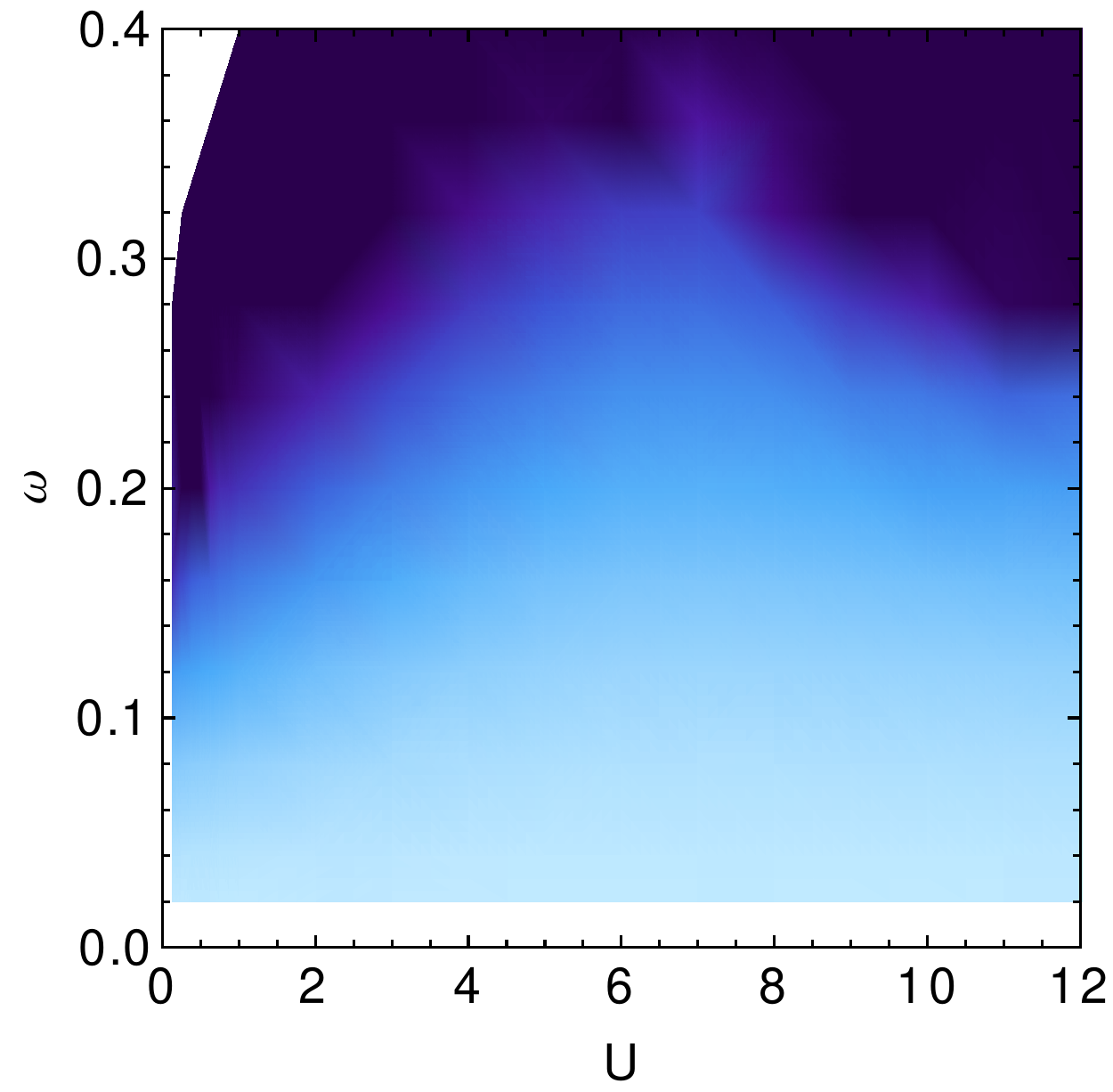} \includegraphics[width=0.1\columnwidth]{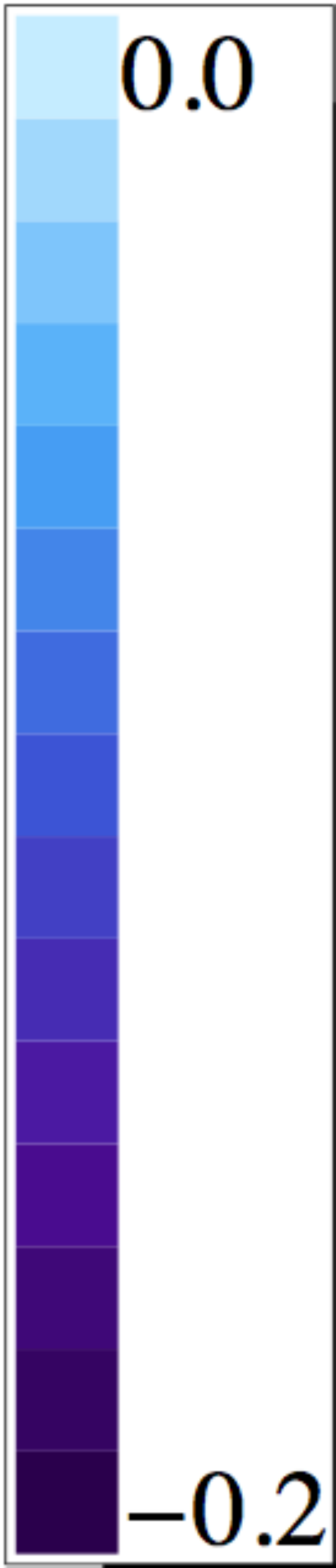}
\caption{Geometric part of the quasienergy $\Omega$ as a function of $(U,\omega_0)$ for $\epsilon=4$ and $\phi_0=0$.}
\label{fig:kinematic} 
\end{figure}

\end{document}